\documentclass[fleqn,usenatbib]{mnras}

\usepackage{newtxtext,newtxmath}

\usepackage[T1]{fontenc}

\DeclareRobustCommand{\VAN}[3]{#2}
\let\VANthebibliography\thebibliography
\def\thebibliography{\DeclareRobustCommand{\VAN}[3]{##3}\VANthebibliography}

\usepackage{graphicx}	
\usepackage{amsmath}	
\usepackage{booktabs}
\usepackage{multirow}
\usepackage{xcolor} 
\usepackage{ulem} 

\DeclareMathOperator{\sinc}{sinc}
\newcommand\chai[1]{{#1}} 
\newcommand\chen[1]{{#1}} 
\newcommand\rmen[1]{{}} 
\newcommand\rg{\text{r}_\text{g}}

\title[GRS~1915+105 phase-resolved QPO]{Phase-resolved spectroscopy of a quasi-periodic oscillation in the black hole X-ray binary GRS~1915+105 with \textit{NICER} and \textit{NuSTAR}}

\author[E. Nathan et al.]{Edward Nathan$^1$, 
Adam Ingram$^\chen{1,2}$,
Jeroen Homan$^\chen{3}$,
Daniela Huppenkothen$^4$,
Phil Uttley$^5$,
\newauthor
Michiel van der Klis$^5$,
Sara Motta$^{2,6}$,
Diego Altamirano$^7$,
Matthew Middleton$^7$
\\
$^1$Department of Physics, Astrophysics, University of Oxford, Denys Wilkinson Building, Keble Road, Oxford, OX1 3RH, UK\\
\chen{$^2$School of Mathematics, Statistics and Physics, Newcastle University, Herschel Building, Newcastle upon Tyne, NE1 7RU, UK}\\
$^3$Eureka Scientific, Inc., 2452 Delmer Street, Oakland, CA 94602, USA\\
$^4$SRON, Netherlands Institute for Space Research, Sorbonnelaan 2, 3584 CA Utrecht, The Netherlands\\
$^5$Anton Pannekoek Institute for Astronomy, University of Amsterdam, Science Park 904, 1098 XH, Amsterdam, The Netherlands\\
$^6$Istituto Nazionale di Astrofisica, Osservatorio Astronomico di Brera, via E. Bianchi 46, I-23807 Merate (LC), Italy\\
$^7$Department of Physics and Astronomy, University of Southampton, Highfield, Southampton, SO17 1BJ, UK
}

\date{Accepted XXX. Received YYY; in original form ZZZ}

\pubyear{2021}

\begin{document}
\label{firstpage}
\pagerange{\pageref{firstpage}--\pageref{lastpage}}
\maketitle

\begin{abstract}
Quasi-periodic oscillations (QPOs) are often present in the X-ray flux from accreting stellar-mass black holes (BHs). If they are due to relativistic (Lense-Thirring) precession of an inner accretion flow which is misaligned with the disc, the iron emission line caused by irradiation of the disc by the inner flow will rock systematically between red and blue shifted during each QPO cycle. 
Here we conduct phase-resolved spectroscopy of a $\sim2.2$~Hz type-C QPO from the BH X-ray binary GRS~1915+105, observed simultaneously with \textit{NICER} and \textit{NuSTAR}. We apply a tomographic model in order to constrain the QPO phase-dependent illumination profile of the disc. We detect \chai{the predicted} QPO phase-dependent shifts of the iron line centroid energy, \chai{with our}
best fit
\chai{featuring}
an asymmetric illumination profile ($>2\sigma$ confidence).
\chai{The observed line energy shifts can alternatively be explained by the spiral density waves of the accretion-ejection instability model. However we}
additionally measure a significant ($>3\sigma$) modulation in reflection fraction,
\chai{strongly favouring}
a geometric QPO origin. We infer that the disc is misaligned with previously observed jet ejections, which is consistent with the model of a truncated disc with an inner precessing hot flow. However \chai{our inferred} disc inner radius is small ($r_\text{in} \sim 1.4~GM/c^2$). 
\chai{For this disc inner radius, Lense-Thirring precession cannot reproduce the observed QPO frequency. In fact, this disc inner radius is incompatible with the predictions of all well-studied QPO models in the literature.}

\end{abstract}




\section{Introduction}
\label{sec:intro}
In a black hole (BH) X-ray binary system (XRB), the BH accretes matter from its stellar companion via a geometrically thin, optically thick accretion disc \citep{Shakura1973, novikov1973} which radiates a multi-temperature blackbody spectrum.  We also see a power law component as a result of photons being Compton up-scattered by a population of hot electrons near the central BH \citep{thorne1975, sunyaev1979}, typically referred to as the \textit{corona}.  This power law component has lower- and upper-cutoffs determined by the temperature of the seed photons and electrons respectively. The third and final major spectral component comes from a fraction of the coronal photons irradiating the disc and being scattered into our line-of-sight.  As a result of being reprocessed in the disc's atmosphere, these photons have a \textit{reflection} spectrum with characteristic features. The scattering produces a broad Compton hump peaking at $\sim20-30$~keV \citep[e.g.][]{Lightman1980}; and there are many spectral lines,
the strongest being the Fe~K$\alpha$ line at $\sim6.4$~keV \citep{george1991, ross2005, Garcia2013}.  These reflection features are distorted and broadened from their rest-frame energies by Doppler shifting and boosting due to the relativistic orbital speed of the disc material, and general relativistic effects due to the strong field gravity around the compact object \citep{Fabian1989}. 
Modelling of these reflection features has been used to trace the inner disc radius, which leads to the BH spin if the disc extends down to the innermost circular stable orbit (ISCO) \cite[e.g.][]{Plant2014,Garcia2015}.

XRBs are usually discovered as transient events, as they undergo outbursts typically lasting weeks to months.  During these outbursts they increase in X-ray flux from the quiescent level by multiple orders of magnitude, but are also seen to transition between different canonical X-ray spectral-timing accretion states. After rising from quiescence, the source is initially in the \textit{hard state}, where the X-ray spectrum is dominated by the power law component. After rising to the peak of the hard state, the source transitions to the disc-dominated \textit{soft state} via the \textit{intermediate state}. Eventually, the source transitions back to the hard state, always at a lower flux than the hard-to-soft transition, and then finally back to quiescence.
The radio properties are correlated with the X-ray state, with a steady jet observed in the hard state and a discrete ejection observed during the hard-to-soft transition \citep[e.g.][]{Done2007review, Fender2004jets, Belloni2010, Fender2012}.

While the physics of the accretion disc is relatively well understood, the structure of the corona is still debated.  
One popular model is the truncated disc model \citep{eardley1975, Ichimaru1977, Done2007review} whereby, in the hard and intermediate states, the disc is truncated at some radius greater than the ISCO of the BH; inside of this truncation radius the flow becomes geometrically thick and optically thin, which is the observed corona. In this model, the truncation radius of the thin disc decreases during the rise from quiescence until it reaches the ISCO in the soft state, before it moves out once again during the decay back to quiescence.
Other models suggest that the corona sits above the disc \citep{galeev1979, haardt1991}, or that the corona is actually outflowing in such a way that is the base of a jet \citep{miyamoto1991, Fender1999, markoff2005}.

Quasi-periodic oscillations (QPOs) are often seen in the light curves of XRBs. These are characterised by a narrow peak with finite width in their power spectra, and are often accompanied by higher harmonics \citep[see e.g. for a review][]{Ingram2019review}.
Here we focus on a `type-C' low-frequency QPO in a BH XRB.  These are seen with fundamental frequency evolving from $\sim 0.1-10$~Hz as the spectral state evolves through the hard and intermediate states \citep{Wijnands1999broadband, vanderKlis2006rapid,Motta2011}. 

Models of low-frequency QPOs in the literature (see \citealt{Ingram2019review} for a discussion) can generally be classified into two types: \textit{intrinsic} -- whereby the intrinsic luminosity of the accretion flow oscillates -- or \textit{geometric} -- whereby the observed oscillation in flux is instead caused by a variation of the beaming pattern of the corona.

Intrinsic models include resonant oscillations in a property of the accretion flow such as accretion rate, pressure, or electron temperature
\citep[e.g.][]{cabanac2010, oneill2011, Karpouzas2021}. For instance, an oscillating shock at the interface between disc and corona (the \textit{propagating oscillatory shock} -- POS -- model: \citealt{Chakrabarti1993}), or spiral density waves in the disc set-up by instabilities in the vertical magnetic field (the \textit{accretion ejection instability -- AEI --} model: \citealt{tagger1999}).

Geometric models mostly focus on relativistic (Lense-Thirring) precession \citep{lense1918}, which is induced in orbits that are not aligned with the BH spin axis by the frame dragging effect. The \textit{relativistic precession model} \citep[RPM:][]{stella1998, stella1999} considers precession frequencies of a test mass in the accretion flow (representing e.g. a hot-spot or over-density). Another example is corrugation modes (c-modes): transverse standing waves in the disc height with resonant angular frequency related to the Lense-Thirring precession frequency \citep[][]{wagoner1999}. \citet{schnittman2006} instead considered a precessing ring in the disc. 
\citet{Ingram2009} proposed that within the truncated disc model the entire corona precesses \citep[as seen in simulations by][]{fragile2007} whereas the disc stays stationary due to viscous diffusion 
\citep{bardeen1975, liska2019}. The precession frequency of the corona is a weighted average over all radii in the corona of the test mass Lense-Thirring precession frequency \citep{Motta2018}.
Alternatively, or additionally, the jet base could be precessing, as has recently been seen in General Relativistic Magneto-hydrodynamic (GRMHD) simulations \citep{liska2018}. 
We note that some of the models classed here as intrinsic also include some geometrical aspect; e.g. an expanding and contracting corona in the POS model and the spiral arms in the AEI model.

\citet{Motta2015} and \citet{Heil2015} showed that higher inclination sources appear to display stronger QPOs, providing strong evidence in favour of a geometrical effect rather than some intrinsic fluctuation in the X-ray luminosity.  Further to this, \citet{vandenEijnden2017} found a possible inclination dependence of QPO phase lags, which also supports a geometrical origin for type-C QPOs. 
It is also known that the power law spectral component varies with much larger RMS than the disc component, indicating an origin in the corona \citep{Sobolewska2006, axelsson2013} as opposed to the disc.
The precessing corona model naturally reproduces these observational properties, and additionally predicts 
that the reflection spectrum is modulated, as the precessing corona illuminates the disc asymmetrically.
The observer sees light coming from different patches of the disc undergoing different boosting and shifting due to differing line of sight velocities of the disc material.
Therefore an asymmetric illumination profile which varies with QPO phase will highlight different patches of the disc at different phases of the QPO cycle, and hence cause the broadening profile of the reflection spectrum to change.  This effect would be seen as a `rocking' of the Fe~K$\alpha$ line, where the profile and centroid energy change over the course of each QPO cycle \citep{Ingram2012,You2020}.

In this paper we study the QPO phase dependence of the reflection spectrum by performing
phase-resolved spectroscopy of a $\sim 2.2$ Hz QPO from the BH XRB GRS~1915+105, using the technique first applied to the same source by \citet{Ingram2015grs1915}.
\citet{Ingram2016} made further improvements to the technique in order to constrain a modulation in the Fe line centroid energy from the BH XRB H~1743-332, and \citet{Ingram2017h1743} introduced a tomographic model.
Likewise, \citet{Stevens2016} presented phase-resolved spectroscopy of GX~339-4, introducing the use of the cross-correlation function.
Here we present further sophistication to the \citet{Ingram2015grs1915} phase-resolving technique, and to the tomographic model. 
In Section~\ref{sec:data} we present details of our observations and data reduction procedure. In Section~\ref{sec:PRS} we lay out the steps of our improved phase-resolving method. Section~\ref{sec:model} contains the details of our tomographic model, and the results of fitting this model to the phase-resolved spectra are presented in Section~\ref{sec:fits}. We discuss our results in Section~\ref{sec:discussion}.

\section{Observations}
\label{sec:data}
\textit{The Neutron star Interior Composition ExploreR} \citep[\textit{NICER};][]{Gendreau2016} and \textit{the Nuclear Spectroscopic Telescope ARray} \citep[\textit{NuSTAR};][]{Harrison2013} observed GRS~1915+105 quasi-simultaneously on $8^\text{th}-9^\text{th}$ June 2018\footnote{MJD 58277-58278}. The details of the observations are summarised in Table~\ref{tab:obs_details}.
In this section, we detail our data reduction procedure and present the basic spectral and timing properties of the data. When analysing the data we make use of \citet{heasoft2014}, and custom code written in Python3~\citep{python3_2009}.

\begin{figure}
    \includegraphics[width=\columnwidth]{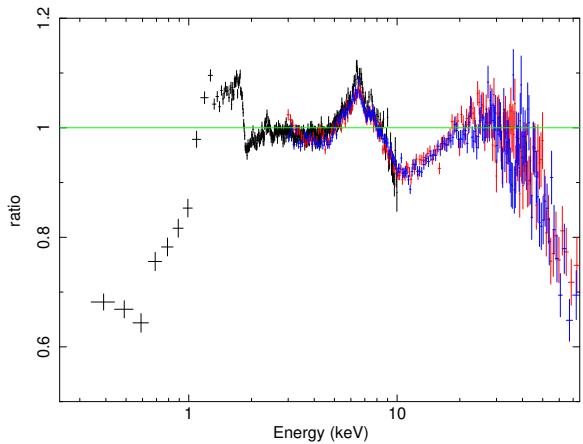}
    \caption{Ratio of the \textit{NICER} $0.24-10$~keV, and the \textit{NuSTAR} FPMA and FMPB $3-75$~keV spectra in black, red, and blue respectively, to an absorbed power law model with photon index 2.01, and absorption with hydrogen column density $6.2\times10^{22}~\text{cm}^{-2}$.}
    \label{fig:ratio}
\end{figure}

\subsection{ \textit{NuSTAR} Data Reduction }
\label{sec:NuSTAR}
We used the \textit{NuSTAR} analysis software, \textsc{\textit{NuSTAR}DAS}~v1.8.0 with \textsc{heasoft}~v6.22. We generated a cleaned event list with associated list of good time intervals (GTIs) for both focal plane modules (FPMs) -- FPMA and FPMB -- using \texttt{nupipeline}. From this we used \texttt{nuproducts} to extract source and background spectra from 49.2 arcsec circular regions and generate spectral response files. We find that the source contributes $>99.9$ per cent of the total counts measured by \textit{NuSTAR}. We did not perform a background subtraction when extracting light curves we use for the timing analysis, as the background is not expected to be variable on the QPO period. We use our own custom code to extract FPMA and FPMB source region light curves from the cleaned event list in 11 broad energy channels in the energy range $3-78$~keV.
We used the \textsc{ftool} \texttt{rbnrmf} to re-bin
the spectral response files into these 11 energy bands.

\subsection{ \textit{NICER} Data Reduction }
\label{sec:NICER}
We used the \textit{NICER} analysis software, \textsc{\textit{NICER}DAS}~v2018-04-13\_V004 with \textsc{heasoft}~v6.24.  We extracted the MPU-merged, uncleaned event lists with the \textsc{ftool} \texttt{\textit{NICER}l2} for each of the two \textit{NICER} observations, which we then merged together using \texttt{nimpumerge} and cleaned with \texttt{\textit{NICER}clean}. 
We used filter and GTI files which were combined from those of the separate obs IDS using \textsc{ftool}s \texttt{ftmerge} and \texttt{nimaketime}.  This results in a single cleaned event list for the two obs IDs combined.

We extracted a flux-energy spectrum from the resulting merged event list using \textsc{xselect}, and estimated the instrumental background
with the \texttt{\textit{NICER}gof.bkg} version 0.5 python script \citep{Remillard2021_preprint}. 
We find that the source contributes 98.6 per cent of the total counts measured by \textit{NICER}. Again, we did not perform a background subtraction when extracting light curves we use for the timing analysis, as the background is not expected to be variable on the QPO period.
We used the spectral response files 
`nixtiref20170601v002.rmf' and `nixtiaveonaxis20170601v004.arf' from \textsc{caldb}.
We extracted light curves from the merged event list in 40 broad energy channels in the energy range $0.3-10$~keV using our own custom code. We used the \textsc{ftool} \texttt{rbnrmf} to re-bin the spectral response files into these 40 energy bands.

\subsection{Energy spectrum}
\label{sec:energy_spec}
Fig.~\ref{fig:ratio} shows the \textit{NICER} (black) and \textit{NuSTAR} (red: FPMA; blue: FPMB) background subtracted flux-energy spectrum plotted as a ratio to a folded absorbed power-law model. We set the hydrogen column density to $N_H=6.2 \times 10^{22}~\text{cm}^{-2}$ (the absorption model is \textsc{tbabs} with the abundances of \citealt{Wilms2000}) and the power-law index to $\Gamma=2.01$ for all three spectra, but allow the three spectra to each have their own normalisation.
We see strong reflection features including an iron line at $\sim 6.4$ keV and a broad Compton hump peaking at $\sim 30$ keV. We also see that the cross-calibration between \textit{NICER} and \textit{NuSTAR} is excellent in the $\sim 3-10$ keV energy range in which their band passes overlap. At energies below $\sim 2.7$ keV, the \textit{NICER} spectrum includes features that are likely due to calibration uncertainties\chen{, and the `shelf' of the response from higher energy photons \citep{NICER_resp} which is dominant below $\sim1$~keV due to the astrophysical absorption leaving very few source photons at low energies}. We therefore only consider energy channels $>2.7$~keV in our spectral analysis.
We note there is a cross-calibration discrepancy between \textit{NuSTAR} FMPA \& FMPB in the energy range $\sim3-3.5$~keV due to a tear in the Multi Layer Insulation around \textit{NuSTAR}'s FMPA \citep{madsen2020}, so we also ignore the FMA energy channels $<3.48$~keV.

\subsection{Power spectrum}

\begin{figure}
    \centering
    \includegraphics{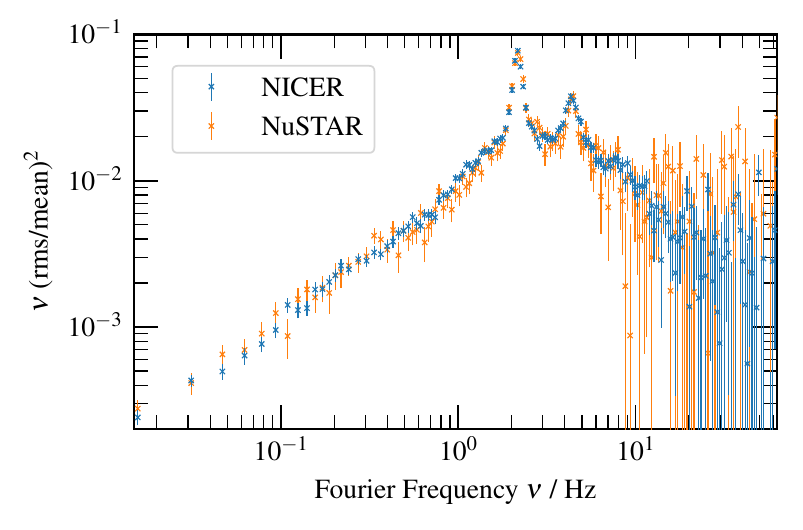}
    \caption{The 3-10~keV power spectra of the \textit{NICER} (blue) and \textit{NuSTAR} (orange) observations.  The \textit{NICER} spectrum is Poisson-noise subtracted, while the \textit{NuSTAR} spectrum is estimated from the co-spectrum between FPMA and FMPB.  The power spectra have been ensemble-averaged and geometrically re-binned to reduce the number of bins by a factor of 25.}
    \label{fig:3_10_keV_powerspectra}
\end{figure}

Fig.~\ref{fig:3_10_keV_powerspectra} shows the 3-10 keV power spectrum calculated for the merged \textit{NICER} observation (blue) and the \textit{NuSTAR} observation (orange). For both observatories, we extract light curves with time step $\delta t$ from the cleaned event list. For the purposes of ensemble averaging \citep[e.g.][]{vanderKlis1989}, we split the light curves into $M$ segments labelled $1 \leq m \leq M$, each having $N$ time bins and therefore a length $T=N {\delta t}$.  Except for when otherwise stated we use $N=8192$ and $\delta t=1/128$~s throughout this paper, so our segments are $T=64$~s long. The \textit{NICER} power spectrum is calculated in the standard way (the magnitude-squared of the fast Fourier transform (FFT) of the lightcurve), with a constant Poisson noise level subtracted \citep{vanderKlis1989, Uttley2014}. For \textit{NuSTAR} we instead calculate the co-spectrum between the FPMA and FPMB \citep{Bachetti2015} in order to avoid instrumental features caused by the fairly large \textit{NuSTAR} deadtime of $t_D \approx 2.5$ ms. We also correct for the suppression of variability caused by the \textit{NuSTAR} dead time using the simple
formula $(\text{RMS}_\text{det}/\text{RMS}_\text{intr}) \approx 1 / ( 1 + t_D r_\text{intr} ) = (r_\text{det}/r_\text{intr})$, where $\text{RMS}_\text{det}$ and $\text{RMS}_\text{intr}$ are respectively the detected and intrinsic RMS variability amplitudes and $r_\text{det}$ and $r_\text{intr}$ are the detected and intrinsic count rates \citep{Bachetti2015}. For this observation, the ratio of detected to intrinsic variability is $\text{RMS}_\text{det}/\text{RMS}_\text{intr} = 0.754$ (recorded in the \textit{NuSTAR} spectral files as the keyword ‘\texttt{DEADC}’). We see that the \textit{NuSTAR} co-spectrum is very similar to the \textit{NICER} power spectrum. In both, we see a strong \chen{low-frequency} QPO with a fundamental frequency of $\sim 2.2$ Hz and a $2^\text{nd}$ harmonic at twice that frequency. \chen{We can classify this low-frequency QPO as `type-C' based upon its frequency,  RMS$\gtrsim15\%$, and the presence of the `flat-top' broadband noise (see e.g. \citealt{Ingram2019review} for a summary of the characteristic features of type-A, B, and C low frequency QPOS).} 

\subsection{Spectral Timing State}

\citet{Belloni2000} identified 15 different states from spectral and variability patterns that GRS~1915+105 transitions between 
\citep[also see][]{huppenkothen2017}, most of which have only to date been observed in one other XRB \citep{altamirano2011}. 
From the flux-energy spectrum and power spectrum, it is clear that GRS~1915+105 was in the designated $\chi-$state during this observation.  
The $\chi-$state is one of the few that behaves similarly to one of the canonical states, and corresponds to the hard state. The $\chi-$state can either be radio loud or quiet; during our X-ray observations the source was radio quiet \citep{motta2021}.
This happens to be the dimmest $\chi-$state ever observed, preceding the transition of the source into its current heavily obscured state \citep{motta2021}. The QPO frequency of $\sim 2.2$ Hz is also somewhat special, since it is at around this QPO frequency when the QPO phase lags transition from positive (hard photons lag soft photons) to negative (soft photons lag hard photons): the phase lag reduces approximately linearly with the log of QPO frequency, passing through zero at $\nu_\text{qpo} \sim 2$ Hz \citep{reig2000,qu2010,vandenEijnden2017,zhang2020}.

\begin{table}
	\centering
	\caption{Details of the simultaneous observations from \textit{NICER} and \textit{NuSTAR} on the $8^\text{th}-9^\text{th}$ June 2018 (MJD 58277-58278). The \textit{NICER} observation was split into two, however these are merged for our analysis.  The net count-rate is reported for the background-subtracted spectra used for the flux-energy fits: $3.5-75$, $3-75$~keV for \textit{NuSTAR}; $2.7-10$~keV for \textit{NICER}.}
	\label{tab:obs_details}

\begin{tabular}{lcccc} 
	\hline
	\multirow{2}{*}{Mission} & \multicolumn{2}{c}{\textit{NuSTAR}} & \multicolumn{2}{c}{ \multirow{2}{*}{\textit{NICER}} }\\
	& FPM~A & FPM~B & \multicolumn{2}{c}{} \\
	\hline
	ObsID & \multicolumn{2}{c}{80401312002} & 1103010157 & 1103010158 \\
	
	Start time & \multicolumn{2}{c}{12:01:09} & 11:42:40 & 23:49:26 \\
	End time & \multicolumn{2}{c}{05:31:09} & 22:55:20 & 05:05:40  \\
	Net count rate / s$^{-1}$ & 60.2 & 59.6 & \multicolumn{2}{c}{196.8}\\
	Exposure time / s & 26166 & 26512 & 15386 & 5033\\
	\hline
\end{tabular}

\end{table}

\section{Phase Resolved Spectroscopy}
\label{sec:PRS}
The aim of phase resolved spectroscopy is to investigate how the energy spectrum of the source varies with QPO phase, a task complicated by the `quasi-' nature of the oscillation which prevents more direct approaches such as phase-folding.  We therefore employ the techniques pioneered by \citet{Ingram2015grs1915} to consider the QPO waveform in different energy bands, considering its phase-average and first two harmonics.  To do this, we extract three key pieces of information:
\begin{itemize}
    \item The amplitude (RMS) of each harmonic in each energy band $\sigma_j(E)$. We find this by fitting an estimate of the power spectrum with a multi-Lorentzian model, as described in Section~\ref{sec:RMS}.
    \item The phase lag of each harmonic in each energy band $\Delta_j(E)$, relative to the phase of the corresponding harmonic in a reference band.  This comes from using the cross spectrum between the lightcurve of the subject energy band, and the reference band, as described in Section~\ref{sec:lags}.
    \item The phase difference $\psi$ between the first two harmonics measured within the reference band.  Using the FFT of the reference lightcurve, this is the difference between the phases in frequency bins containing the two harmonics. We use the bi-spectrum to calculate this, which is described in Section~\ref{sec:bispec}.
\end{itemize}

Following \citet{Ingram2015grs1915} and \citet{Ingram2016}, we consider that the count rate $w(E,\gamma)$ in each energy bin denoted by $E$ varies with QPO phase $\gamma$ as
\begin{equation}
    w(E, \gamma) = \mu(E)\left[1 + \sum_{j=1}^{J} \sigma_j(E) \cos\left(j\gamma - \Phi_j(E)\right) \right]\,,
    \label{eqn:average_waveform}
\end{equation}
where $\mu(E)$ is the average count rate in the energy band, $\sigma_j(E)$ is the average RMS of the $j^\text{th}$ QPO harmonic\footnote{The average RMS is often reported as $\langle\sigma_j(E)\rangle$, with an additional multiple of $\sqrt{2}$ included in Eq.~\ref{eqn:average_waveform}. This is to make explicit that the average RMS of a sine-wave is $1/\sqrt{2}$.  For simplicity we neglect this, keeping in mind our model is normalised to expect the average RMS.}, and $\Phi_j(E)$ is the phase offset of the $j^\text{th}$ QPO harmonic.  The phase-offset is split into an energy-dependent phase-lag $\Delta_j(E)$, which is the phase lag of a harmonic compared to the same harmonic in the reference band light curve, and the phase difference $\psi$ between the two harmonics in the same reference band light curve.  These are combined so that
\begin{equation}
    \begin{split}
        \Phi_1(E) &= \Phi_1 + \Delta_1(E) \\
        \Phi_2(E) &= 2(\Phi_1 +\psi) + \Delta_2(E),
    \end{split}
    \label{eqn:phase_offsets}
\end{equation}
where, following \citet{Ingram2015grs1915}, we choose to set the arbitrary phase of the first harmonic to $\Phi_1=\pi/2$. 

Putting this together, for $j\geq1$ we get the Fourier transformed (FT) spectra \citep{Ingram2016}
\begin{equation}
    W_j(E) = \mu(E) \sigma_j(E) \text{e}^{\text{i}\Phi_j(E)},
    \label{eqn:FT_spectra}
\end{equation}
plus the phase-average $W_0(E)=\mu(E)$, which is trivially the flux-energy spectrum. We fit the theoretical model described in the following section simultaneously to the real and imaginary parts of the $j=1$ and $j=2$ FT spectra, and the flux-energy spectrum ($j=0$). We use the full spectral resolution of the instrument for the flux-energy spectrum (grouped to have $\ge 30$ counts in each energy channel) and subtract background. For the $j\ge 1$ QPO harmonics, we instead use the broader energy bands defined in Sections \ref{sec:NuSTAR} and \ref{sec:NICER} and do not perform a background subtraction. This treatment of the background is appropriate because the background is not expected to be variable on the QPO period.

\chen{The following subsections which describe these parts of the analysis are each further split into two sub-subsections, with the first describing the method used, and the second presenting the results obtained from the observations analysed in this paper.}

\subsection{QPO frequency tracking}
\label{sec:freq_track}

\begin{figure*}
    \centering
    \includegraphics[]{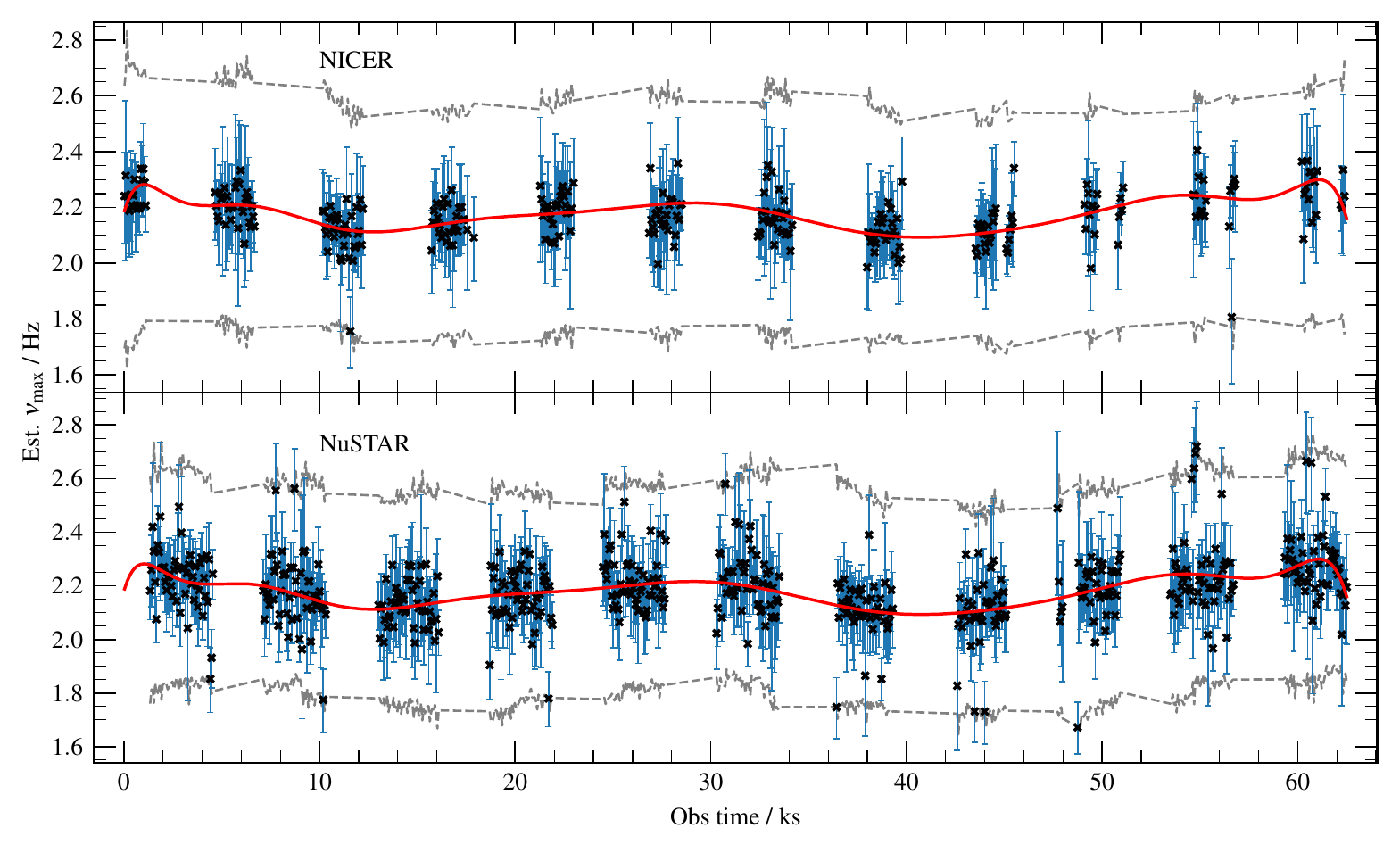}
    \caption{The estimated value of the QPO frequency $\nu_\text{max}$ measured in each 64~s long segment of the two light curves, with the error bars denoting the estimated FWHM.  The grey dashed line show the dynamic bounds for the QPO tracking algorithm.  The red line is a $15\deg$ polynomial simultaneously fit to the estimated QPO frequencies, which we then use to get the QPO frequency at arbitrary time.}
    \label{fig:qpo_tracking}
\end{figure*}

During the observations, the frequency of the QPO drifts over time by $\sim\pm5\%$.  To account for this, we identify the frequency of the QPO during each of the M segments so that we can later average Fourier products over the frequency range containing the instantaneous QPO frequency.  

\subsubsection{\chen{Method}}

\chai{We determine the QPO frequency for each segment by fitting a model to the power spectrum of each segment of the \textit{NICER} and \textit{NuSTAR} (we sum the FPMA and FPMB counts) light curves. 
Our model comprises of four Lorentzian functions \citep{vanStraaten2002}, two of which are harmonically locked to represent the two QPO harmonics, and a Poisson noise component. The Poisson noise component is very simple for \textit{NICER}, taking a constant value of $2/\mu$ where $\mu$ is the mean count rate \citep{vanderKlis1989}. The Poisson noise is much more complicated for \textit{NuSTAR}, due to the large detector deadtime, $t_D$. Since there is currently no accurate deadtime model for \textit{NuSTAR}, we model the Poisson noise with the function \citep{Bult2017}
\begin{equation}
    f\left(\nu|A,B,t_D\right)=A-2 B t_{D}\sinc{\left(2\pi\nu t_{D}\right)}\,,
    \label{eqn:sinc_deadtime_model}
\end{equation}
and determine the parameters $A$, $B$ and $t_D$ by fitting the above function to the power spectrum averaged over the entire observation. We only consider $\nu > 20$ Hz since this frequency range is Poisson noise dominated. This fit yields $A=(1.783\pm0.006)\times10^{-2}$, $B=0.15\pm0.01$ and
$t_D=3.3\pm0.2~\textrm{ms}$. We then freeze $A$, $B$ and $t_D$ to these best fitting values for the remainder of the analysis.}

\chen{
As each frequency bin of a single power spectrum (without any ensemble or frequency averaging) follows a $\chi^2$ probability distribution with two degrees of freedom\footnote{Apart from the bin at the Nyquist frequency which is a $\chi^2$ distribution with only a single degree of freedom, but this bin was ignored in our calculations for simplicity}~\citep{vanderKlis1989}, we are unable to use $\chi^2$ minimisation for the model fit as this depends on each frequency bin following a Gaussian distribution. We therefore use the maximum likelihood estimation method described in \citet{Barret2012}. This method does not require the probability distribution to be Gaussian, it only requires it to be known analytically. We find best fitting model parameters, including the frequency of the QPO fundamental, by maximising the likelihood function calculated assuming a $\chi^2$ probability distribution with two degrees of freedom.

It is important to note that the probability distribution underlying each frequency bin of a un-averaged co-spectrum is not known analytically \citep{huppenkothen2018}. We are therefore unable to use the maximum likelihood method on the co-spectrum between FPMA and FPMB, which is why we resort to modelling the Poisson noise of the power spectrum, which does have well-understood statistics.
}

\subsubsection{\chen{Results}}

Fig.~\ref{fig:qpo_tracking} shows the resulting measurements of QPO frequency \chen{(black crosses) } as a function of time for \textit{NuSTAR} (top) and \textit{NICER} (bottom). Rather than uncertainties, the error bars shown are the measured full width at half maximum (FWHM) of the Lorentzian function representing the QPO fundamental. In our fits, we restricted the QPO frequency to a range given by the running average of the QPO frequencies of the previous 5 segments $\pm$ $3/2$ times the running average FWHM of the previous 5 segments, starting with the QPO frequency and FWHM from the average power spectrum. \chai{This range is represented by the grey dashed lines}. To smooth out the results of our tracking algorithm\footnote{The difference in the scatter of QPO frequencies between the lower count rate \textit{NuSTAR} and higher count rate \textit{NICER} data suggests that this is noise in the measurement rather than intrinsic short timescale changes.} we use a degree-15 polynomial to model QPO frequency with time, that we fit simultaneously to the \textit{NICER} and \textit{NuSTAR} data (the solid red line in Fig.~\ref{fig:qpo_tracking}). \chen{It is encouraging that the instantaneous QPO frequencies we measure here are very similar to those inferred by \citet{huppenkothen2021} using a more sophisticated method (see their fig. 23)}

\subsection{Phase lag spectrum}
\label{sec:lags}

\subsubsection{\chen{Method}}
In order to calculate the energy-dependent phase lag of the $j^\text{th}$ harmonic $\Delta_j(E)$, we first calculate the cross-spectrum between the light curve of the energy band centred on energy $E$ (the subject band) and the light curve summed over all energy channels (the reference band) for each of the $M$ segments. For the $m^\text{th}$ segment, the cross-spectrum as a function of frequency is
\begin{equation}
    G_m(\nu, E) \propto S_m(\nu,E) R_m^*(\nu) 
    \label{eqn:cross_spectrum}
\end{equation}
where $R_m(\nu)$ and $S_m(\nu, E)$ are the Fourier transforms\footnote{As we are using the FFT, we actually use discrete frequency bins $\nu_k=k/T$.} of the $m^\text{th}$ segment of the reference and subject band light curves respectively, and the constant of proportionality is a normalisation into fractional RMS. For \textit{NICER}, the photons in the subject band light curve are also in the reference band light curve (since the reference band consists of all the photons detected by \textit{NICER}). This contributes Poisson noise, which we subtract off following \citet{Ingram2019error}.

For \textit{NuSTAR}, we instead extract the subject band light curves from the FPMB and the reference band light curve from the FPMA to ensure that the subject and reference band signals are statistically independent of one another, and therefore the cross-spectrum contains no contribution from deadtime affected Poisson noise. 

We consider the `shifted-and-added' cross-spectrum of the $j^\text{th}$ QPO harmonic by first averaging over the QPO harmonic in each time segment based upon the tracked QPO frequency.  For the $m^\text{th}$ segment we average over the frequency range $\nu = j \nu_\text{qpo}(m) [1 \pm 1/(2Q)]$, where we assume $Q=8$ for the quality factor (a typical value for a type-C QPO; \citealt{Ingram2019review}), using the smoothed estimate from our QPO tracking algorithm for $\nu_\text{qpo}(m)$.  We then average over the time segments to find the overall average value for cross-spectrum for each of the QPO harmonics.

The phase lag for the $j^\text{th}$ harmonic, $\Delta_j(E)$, is the argument of the averaged cross-spectrum of the corresponding harmonic, of which we estimate the uncertainties using the formula from \citet[eq 19]{Ingram2019error}. 

\subsubsection{\chen{Results}}
\label{sec:phase_lag_spectrum_results}
\chen{We display the measured phase lag spectrum} in Fig.~\ref{fig:both_RMS_spectra} (bottom) for \textit{NICER} (blue) and \textit{NuSTAR} (orange). We see that the phase lag of the QPO fundamental is almost constant with energy, which is consistent with previous \textit{RXTE} observations showing that the phase lag monotonically reduces from hard lags (a positive lag vs energy gradient) to soft lags (a negative lag vs energy gradient) as the QPO frequency increases, with the cross over occurring for $\nu_\text{qpo} \sim 2$~Hz \citep[e.g.][]{vandenEijnden2017}. Note that a slight offset between \textit{NICER} and \textit{NuSTAR} lags results from a phase lag between the \textit{NICER} and \textit{NuSTAR} reference bands. Although the offset is very small for this observation because the energy dependence of the lag is subtle, we account for it in our modelling with a floating phase offset.

\subsection{Fractional RMS Spectrum}
\label{sec:RMS}
\subsubsection{\chen{Method}}
To calculate the fractional RMS of the QPO harmonics in each energy band, instead of the power spectrum in that energy band, we boost signal to noise by using the shift-and-added cross-spectrum $G(\nu, E)=\frac{1}{M}\sum_m^M{G_m(\nu+\delta\nu_m,E)}$, where $\delta\nu_m$ is the difference between the QPO frequency in that segment (from our smoothed tracking) and the average QPO frequency.

Since the QPO harmonics are well correlated between energy bands, we assume unity coherence between the subject bands and the reference band, in which case the shifted-and-added power spectrum of the subject band can be written as \citep{Wilkinson2009, Ingram2016}
\begin{equation}
    P_\text{s}(\nu, E) = \frac{|G(\nu,E)|^2 -\hat{b}^2(\nu, E)}{P_\text{r}(\nu)}\,,
\end{equation}
where $\hat{b}^2(\nu, E)$ results from a positive-bias in the calculation of $|G(\nu, E)|^2$ (see \citealt{Ingram2019error} for details, where $b$ is used instead of $\hat{b}$) and  $P_\text{r}(\nu)$ is the Poisson noise subtracted shift-and-added power spectrum of the reference band for \textit{NICER} and the shift-and-added co-spectrum between the FPMA and FPMB for \textit{NuSTAR}. 

We fit a multi-Lorentzian model to the resulting power spectral estimate for each energy band. We use three Lorentzian functions: one with $Q=0$ to represent the broad band noise; and the other two with centroid frequencies and $Q$ tied (such that the centroid frequencies are harmonically related) in order to represent the two QPO harmonics. We normalise our power spectral estimate \citep{belloni1990} and Lorentzian functions \citep{vanStraaten2002} such that the best-fitting normalisation of the two Lorentzian components representing the QPO harmonics gives their fractional RMS. We calculate $1\sigma$ uncertainties on the RMS by searching parameter space for a marginalised $\Delta \chi^2 =1$.

\subsubsection{\chen{Results}}
\label{sec:RMS_spectrum_results}
\begin{figure*}
 \centering
\includegraphics[]{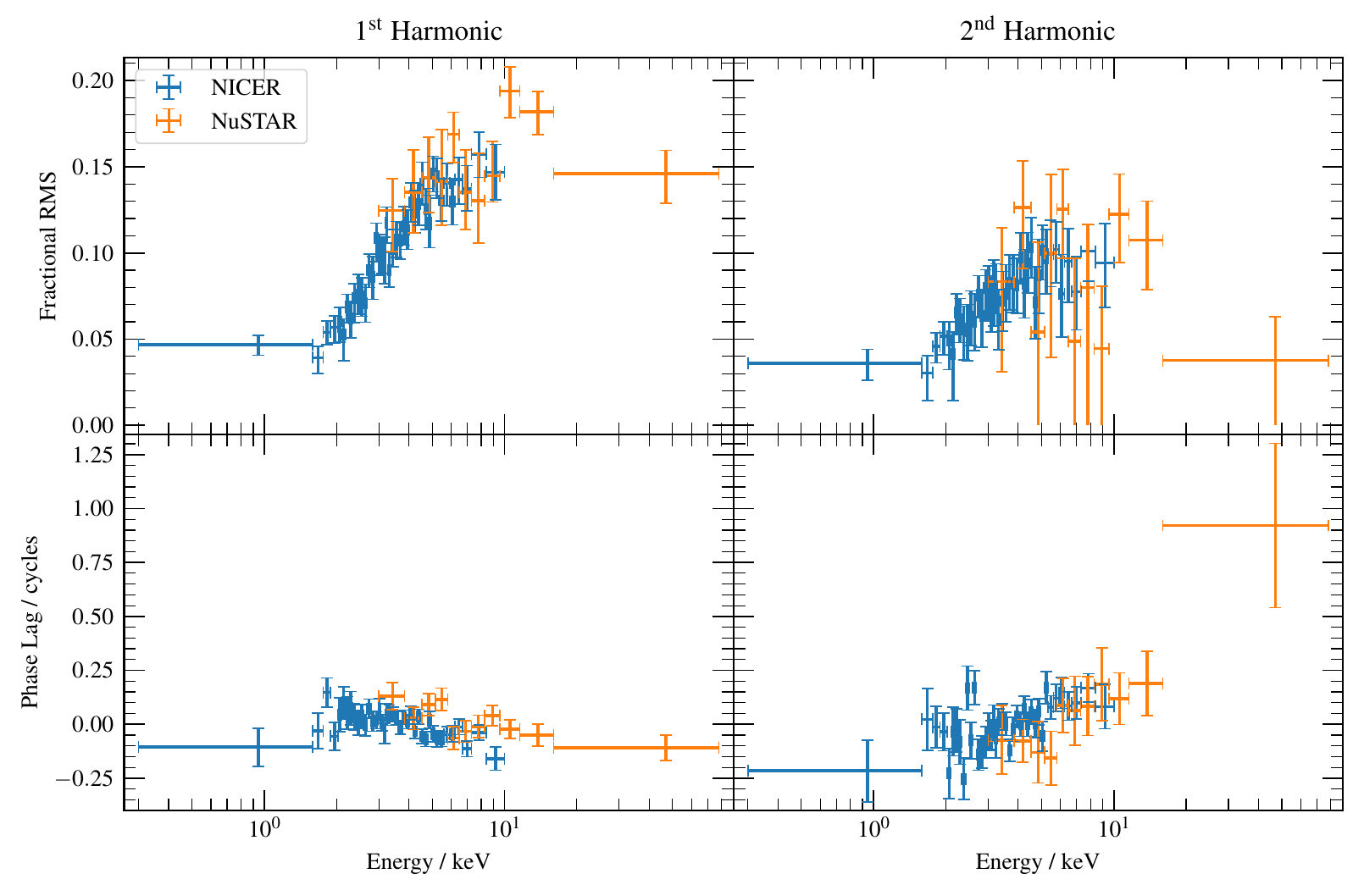}
\caption{Top: The fractional RMS of the first and second harmonics of the QPO in both the \textit{NICER} and \textit{NuSTAR} observations in different energy bands, found by fitting lorenztian functions to the power-spectrum of each energy band, which was estimated from a cross-spectrum between a broad reference energy band and a specific energy band, as described in the text.  As the fractional RMS must be positive points that have $\Delta\chi^2<1$ at zero are shown without an errorbar cap. The fractional RMS is slightly diluted by background photons in the very highest \textit{NICER} energy bands.
Bottom: The phase lag of the first and second harmonics of the QPO, measured against a reference band.  For \textit{NICER} this reference band is the full energy lightcurve, while for \textit{NuSTAR} it is the full energy lightcurve of FMPB.
}
\label{fig:both_RMS_spectra}
\end{figure*}

\chen{We display the measured RMS spectrum} in Fig.~\ref{fig:both_RMS_spectra} (top). Error bars without a lower cap correspond to points consistent with zero within $1\sigma$. We see that, as is typically the case for type-C QPOs, the fractional RMS increases with energy for $E\lesssim 10$ keV before levelling off. We note good agreement between \textit{NICER} and \textit{NuSTAR}.

\subsection{Phase difference between harmonics}
\label{sec:bispec}
\begin{figure*}
    \centering
    \includegraphics[]{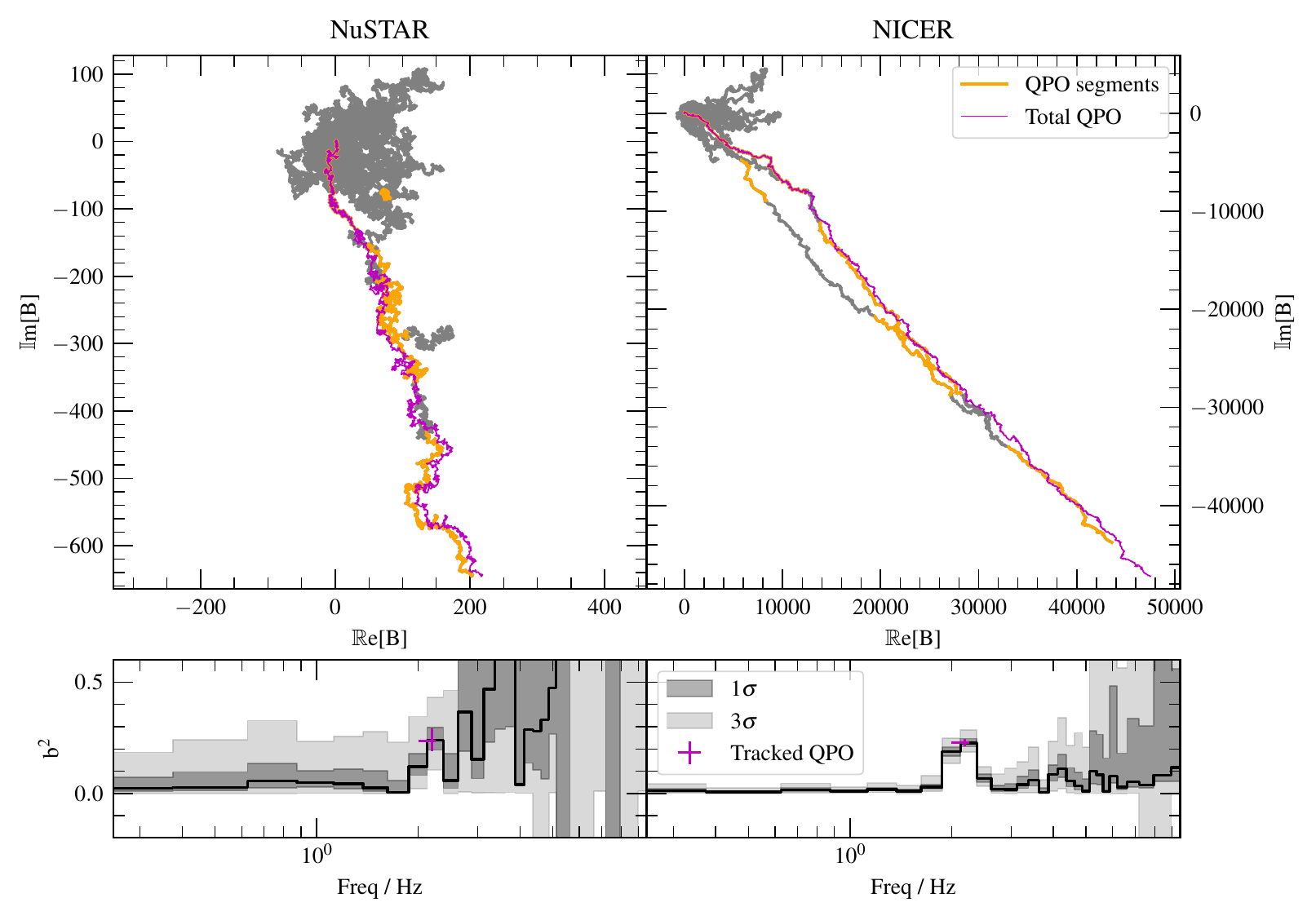}
    \caption{The random walk of the auto-bispectrum $B(\nu_k)$ for frequency bins $\nu_k$ which cover the frequency range $\nu_k<10$~Hz.  Between 6-10~Hz the frequency bins are geometrically re-binned to reduce the number of bins by a factor of 4 between 6-10~Hz.  Each grey chain has steps $B_m(\nu_k)$, but the segments that correspond to the instantaneous QPO frequency are highlighted in orange.  All the segments that contain the QPO are summed into a purple chain.  On the lower panel, the auto-bicoherence $b^2(\nu_k)$ of each chain is given for each frequency. The grey uncertainties, and also the vertical $1\sigma$ uncertainty of the tracked QPO auto-bicoherence, are calculated from percentiles of a bootstrapped population.}
    \label{fig:jellyfish}
\end{figure*}

\subsubsection{\chen{Method}}

As we have the phase lag of each QPO harmonic in the energy bands compared to their counterpart in the reference band, we now need to find the phase lag between the harmonics in the reference band $\psi$.
For this, we use the bispectrum\footnote{We note this method can find the phase-difference between any doublet harmonics, e.g. also between the $2^\text{nd}$ and $4^\text{th}$ harmonics.  In much the same way, higher order polyspectra could be used to find the phase difference between other harmonics, e.g. the trispectrum could be used to find the phase difference between the $1^\text{st}$ and $3^\text{rd}$ harmonics.} as suggested by \citet{Arur2019}. This yields similar results to the method of \citet{Ingram2015grs1915} used in \citet{Ingram2016}, \citet{Ingram2017h1743} and \citet{deRuiter2019}, but is more statistically robust. In particular there is no Poisson noise correction in the \citet{Ingram2015grs1915} method, whereas the effect of Poisson noise on the bispectrum is well covered in the literature \citep[e.g.][]{Wirnitzer1985, vanderKlis1989, kovach2018}. The bispectrum method should therefore be more robust to low count rate observations. The bispectrum is defined as a function of two frequencies, $\nu_1$ and $\nu_2$, such that the bispectrum of the reference band light curve in the absence of Poisson noise is
\begin{equation}
    \mathcal{B}(\nu_1,\nu_2) = \frac{1}{M} \sum_{m=1}^{M} R_m(\nu_1) R_m(\nu_2) R_m^*(\nu_1+\nu_1).
\end{equation}
The phase-difference between harmonics can be retrieved from the `auto-bispectrum', $B(\nu)=\mathcal{B}(\nu_1=\nu,\nu_2=\nu)$, which is
\begin{equation}
    B(\nu) = \frac{1}{M} \sum_{m=1}^{M} B_m(\nu) =\frac{1}{M} \sum_{m=1}^{M} R_m(\nu) R_m(\nu) R_m^*(2\nu).
\end{equation}
Since $R_m(\nu) = |R_m(\nu)|\text{e}^{i\Phi_m(\nu)}$, we see that
\begin{equation}
    B(\nu) = \frac{1}{M} \sum_{m=1}^{M} |R_m(\nu)|^2~|R_m(2\nu)|~\text{e}^{ i [2\Phi_m(\nu) - \Phi_m(2\nu)]}.
\end{equation}
Therefore the bi-phase $\equiv \text{arg}[B(\nu)]=2\Phi(\nu)-\Phi(2\nu)=-2\psi(\nu)$, where we take the phase difference between frequency $\nu$ and frequency $2\nu$ to be $\psi(\nu) = \frac{1}{2}\Phi(2\nu) - \Phi(\nu)$ following \citet{Ingram2015grs1915}. The phase difference between the two QPO harmonics is therefore $\psi = -\frac{1}{2}\text{arg}[B(\nu_\text{qpo})]$.

In order to calculate $B(\nu_\text{qpo})$, we adopt the same shift-and-add technique for the auto-bispectrum as described in the previous section for the cross-spectrum, again employing our smoothed estimate for the instantaneous QPO frequency from Fig.~\ref{fig:qpo_tracking}. The QPO is only coherent on timescales of $\sim Q$~cycles \citep[e.g.][]{vandenEijnden2016}], and so it is reasonable to use segments of duration $T \sim Q/ \nu_\text{qpo}$ \citep{Ingram2015grs1915}. As the QPO frequency in our observations is $\sim2.2$~Hz, and with an assumed quality factor $Q=8$, we use $4$~s long segments.  Using $\delta t=1/128$~s, this gives segments with $N=512$ time bins. We correct the \textit{NICER} auto-bispectrum for Poisson noise as described in Appendix~\ref{app:bispec_poisson_noise}. For \textit{NuSTAR}, we avoid deadtime effects by using FPMA data for $R(\nu)$ and FPMB data for $R(2\nu)$. 

\subsubsection{\chen{Results}}
\label{sec:phase_difference_results}
\chen{
Using the bi-spectrum method described above, we measure $\psi/\pi=0.20\pm0.02$ for \textit{NuSTAR} and $\psi/\pi=0.125\pm0.006$ for \textit{NICER}.
The slight difference in $\psi$ between observatories is statistically significant, and indicates that the QPO waveform depends on photon energy.}

We compare the results using this method to the method used in the literature \citep[e.g.][]{Ingram2016} where $\Phi(\nu)$ and $\Phi(2\nu)$ are taken directly from the FFT, and a minimisation is used to find $\psi$. For \textit{NuSTAR} this gives $\psi/\pi=0.21\pm0.04$, and for \textit{NICER} this gives $\psi/\pi=0.12\pm0.01$, which are consistent with the values from our updated method. \chai{Here, we again avoid the deadtime affected \textit{NuSTAR} Poisson noise by taking $\Phi(\nu)$ from the FFT of the FMPA light curve, but $\Phi(2\nu)$ from the FFT of the FMPB light curve.}

\chai{As discussed in \cite{Ingram2015grs1915}, it only makes sense to measure the phase difference between harmonics if the phases of those two harmonics are correlated. In such a case, the QPO has some well-defined underlying waveform, and it makes sense to do phase-resolved spectroscopy. Otherwise, the spectrum does not vary in shape in a systematic way with QPO phase, and a QPO phase-resolved analysis would thus be meaningless. It is therefore important to measure how well correlated the two dominant QPO harmonics are before continuing. We explore this in the following sub-section.}

\subsubsection{Phase correlation between QPO harmonics}

The auto-bispectrum can also be used to measure the extent to which the phases of the two harmonics are correlated. Specifically, the auto-bicoherence is the modulus of the auto-bispectrum, re-normalised in some useful way. We use the \citet{Kim1979} normalisation for which the auto-bicoherence, $b^2(\nu)$, is unity if the phases of the $\nu$ and $2\nu$ components are perfectly correlated. In the opposite case of a completely uncorrelated signal, $b^2(\nu) \rightarrow 0$ as the number of light curve segments used to calculate $b^2$ tends to $M \rightarrow \infty$, and it becomes meaningless to measure the phase difference between the harmonics at $\nu$ and $2\nu$.

\chen{In the absence of Poisson noise the auto-bicoherence is given by \citep{Kim1979}}
\begin{equation}
    b^2(\nu) = \frac{ \left\vert\sum_{m=1}^M B_m(\nu)\right\vert^2 }{ \sum_{m=1}^M |R_m(\nu)R_m(\nu)|^2 ~ \sum_{m=1}^M |R_m(2\nu)|^2 }.
    \label{eqn:squared_bicoherence}
\end{equation}
We describe how we account for Poisson noise and deadtime effects in the denominator of the above equation in Appendix~\ref{app:bispec_poisson_noise}\chen{, and also describe a bootstrapping technique \citep[following][]{Stevens2016} we use to calculate the errors on the auto-bicoherence in Appendix~\ref{app:bootstrap}.}

\chai{The auto-bicoherence as a function of frequency is shown in Fig.~\ref{fig:jellyfish} (bottom) for \textit{NuSTAR} (left) and \textit{NICER} (right). The black stepped lines show the measured values and the shaded regions represent the $1$ and $3$ $\sigma$ confidence regions (calculated using a bootstrapping method). We see that the auto-bicoherence is consistent with zero for all frequencies except for around the QPO fundamental frequency. This indicates that the phase of the first harmonic is well correlated with that of the second harmonic, and that there is therefore an underlying QPO waveform \citep{Ingram2015grs1915,deRuiter2019}. In contrast, all other pairs of frequencies are uncorrelated.

The width of the feature around the QPO frequency in these plots of the bispectrum could be because the QPO frequency drifts during the observation, and therefore the frequency bin that contains the QPO frequency may not be the same for each segment. We therefore calculate $b^2(\nu_{\rm qpo})$ using the same shift-and-add techniques as described in the previous sub-sections. The result is marked in magenta. The vertical error bar is $1\sigma$ and again calculated by bootstrapping (as discussed in Appendix~\ref{app:bootstrap}), and the horizontal error bar shows the frequency range covered by the QPO fundamental during the observation. We see that this new shifted-and-added bicoherence is consistent with the black stepped line, and therefore the shift-and-add does not enhance the coherence of the QPO for this particular observation. While the coherence between the QPO harmonics is statistically non-zero, it is still not unity. This can be partly explained by the presence of uncorrelated broad band noise at the QPO frequency. However, the QPO dominates over the broad band noise in the power spectrum for $\nu \approx \nu_{\rm qpo}$. It is therefore likely that $\psi$ is not constant in time -- as would be the case for a perfectly periodic oscillation -- but instead varies around a well-defined mean value.}

\chen{We visualise the auto-bispectrum in Fig.~\ref{fig:jellyfish} with `jellyfish plots' (top panels). Here, for each Fourier frequency, we plot $B(\nu)$ as a vector sum on the complex plane. The vector sum for each frequency is plotted as a grey line (and there are a total of 29 frequencies after geometric re-binning above 6~Hz). We see that this forms a random walk on the complex plane. For most frequencies, this random walk forms a blob that never gets far from the origin, indicating that the phase at $\nu$ is poorly correlated with that at $2\nu$. For a narrow range of frequencies, the random walk instead forms a much straighter path that extends far from the origin, indicating good correlation between $\nu$ and $2\nu$. The auto-bicoherence is a measure of how far from the origin the vector sum ends up, and the biphase is a measure of the orientation on the complex plane of the summed vector. The jellyfish plots therefore demonstrate that $b^2(\nu)$ is large for $\nu \approx \nu_{\rm qpo}$ because each segment has a similar bi-phase, and so the segments all line up well on the complex plane.}

\chai{We also demonstrate how the drifting of the QPO frequency during the observation leads to the frequency bin that contains the QPO changing from one segment to another. To do this, we 
mark segments on the jellyfish plots that contain the instantaneous QPO frequency by colouring them orange. We see that the QPO frequency does indeed jump from one frequency bin to another. This is particularly noticeable in the \textit{NICER} jellyfish plot. The magenta line instead shows the vector sum of $B(\nu_\text{qpo})$ that we use to calculate $b^2(\nu_{\rm qpo})$; i.e. we take only the segments that contain the instantaneous QPO frequency and add them on the complex plane. Consistent with the bicoherence plots, we see that the magenta line reaches a comparable distance from the origin to the two grey lines around the QPO frequency.}

\chen{

\subsection{Reconstructed Fourier Transformed Spectra}
\label{sec:reconstructed_FT_spectra}
We use the phase lag spectra $\Delta_1(E)$ and $\Delta_2(E)$ found in Section~\ref{sec:phase_lag_spectrum_results}, the RMS spectra $\sigma_1(E)$ and $\sigma_2(E)$ found in Section~\ref{sec:RMS_spectrum_results}, and the phase difference between harmonics $\phi$ found in Section~\ref{sec:phase_difference_results} to calculate our Fourier Transformed spectra $W_1(E)$ and $W_2(E)$ using Eqs.~\ref{eqn:phase_offsets} and \ref{eqn:FT_spectra}.  As $W_j(E)$ ($j=1,2$) is a complex quantity, we separate these into real and imaginary parts $\Re\left[W_j(E)\right]$ and $\Im\left[W_j(E)\right]$.  Therefore, for both of our \textit{NuSTAR} and \textit{NICER} observations we have calculated 4 spectra
\begin{enumerate}
    \item $\Re\left[W_1(E)\right]$, the real part of the FT spectra of first QPO harmonic,
    \item $\Im\left[W_1(E)\right]$, the imaginary part of the FT spectra of first QPO harmonic,
    \item $\Re\left[W_2(E)\right]$, the real part of the FT spectra of second QPO harmonic,
    \item $\Im\left[W_2(E)\right]$, the imaginary part of the FT spectra of second QPO harmonic,
\end{enumerate}
giving a total of 8 FT spectra.  We also have the phase-average flux-energy spectra `$W_0(E)$' for each of \textit{NuSTAR}'s FMPA and FMPB, plus \textit{NICER}'s, bringing our total to 11 spectra which we will simultaneously fit with the model described in the following section.}

\chai{The resulting spectra are shown in Fig \ref{fig:auto_temp_model_fits}. The left panel shows the phase-averaged flux-energy spectrum, $W_0(E)$, observed by \textit{NICER} (black), FPMA (red) and FPMB (blue). The derived FT spectra are plotted on the right hand side as }\chen{grey and black points. 
The first and second harmonic (i.e. $j=1$ and $j=2$, or in other words the fundamental and the first overtone) are plotted respectively in the first and second panels from the top (as labelled), and grey and black points correspond respectively to the real ($\Re [W_j(E)]$) and imaginary ($\Im [W_j(E)]$) parts respectively.}
\chai{ Open circles correspond to \textit{NuSTAR} and the points with no marker to \textit{NICER}. Note that each part of each harmonic has only one \textit{NuSTAR} FT spectrum, and not one for the FPMA and another for the FPMB. This is because the \textit{NuSTAR} FT spectra are derived by extracting the subject bands of the cross-spectrum from the FPMB and the reference band from the FPMA. Both FPMs are therefore used for this single measurement.}

\section{Theoretical Model}
\label{sec:model}
Our model for the QPO FT calculates the X-ray spectrum as a function of QPO phase, $\gamma$, before Fourier transforming to output the real and imaginary parts of the QPO FT for the zeroth, first and second harmonics. The model is similar to the one described in \citet{Ingram2017h1743}, but with some extra features. As in \citet{Ingram2017h1743}, we assume that the accretion flow has two components: a thin accretion disc and a corona. We assume that the disc is stationary with inner and outer radii $r_\text{in}$ and $r_\text{out}$. We make no assumptions about the shape of the corona, but we assume that its intrinsic bolometric luminosity is constant in time, and variation in the observed flux is caused by us viewing the corona from different directions at different QPO phases. 

The spectrum from the corona is described in Section~\ref{sec:model_corona}. Section \ref{sec:disc_summary} describes the spectrum from the disc, which we split into a thermal component (Section \ref{sec:thermal}) and a non-thermal component (Section \ref{sec:nonthermal}).

\subsection{Corona}
\label{sec:model_corona}
We represent the corona spectrum with the model \textsc{nthcomp} \citep{Zdziarski1996}, which is a power-law (index $\Gamma$ -- such that the \textit{photons} emitted per unit energy is $\propto E^{-\Gamma}$) between low and high energy cut-offs that are respectively governed by the seed photon temperature $T_\text{bb}$ and the electron temperature $T_\text{e}$. We parameterise the QPO phase-dependent bolometric flux observed from the corona as
\begin{equation}
    N_\text{c}(\gamma) = N_0 + A_{1N} \sin[ \gamma - \phi_{1N} ] + A_{2N} \sin[ 2(\gamma - \phi_{2N}) ],
    \label{eqn:Nc}
\end{equation}
where $N_0$, $A_{1N}$, $A_{2N}$, $\phi_{1N}$ and $\phi_{2N}$ are left as free parameters. We parameterise the photon index $\Gamma(\gamma)$ and electron temperature $T_e(\gamma)$ in a similar way (see e.g. Eq.~5 of \citealt{Ingram2017h1743}). We tie $T_\text{bb}$ to the peak disc temperature, which we discuss at the end of Section~\ref{sec:thermal}.

\subsection{Disc}
\label{sec:disc_summary}
The corona irradiates the disc, and since the disc is very optically thick, all of the irradiating flux is reprocessed in the disc atmosphere and re-emitted. The disc is also heated by viscous dissipation of gravitational potential energy, generating intrinsic disc flux. The total radiated flux is the sum of intrinsic and reprocessed flux. We assume that all of the intrinsic flux plus some fraction of the reprocessed flux is in thermal equilibrium with the disc, thus contributing a blackbody component to the emitted spectrum. The total restframe specific intensity emergent from the disc coordinate $(r,\phi)$ at QPO phase $\gamma$ is therefore the sum of this blackbody component, and a non-thermal component which includes well-known `reflection' features such as the iron line and Compton hump
\begin{equation}
    I(E_d,r,\phi,\gamma) = I_\text{bb}(E_d,r,\phi,\gamma) + I_\text{nt}(E_d,r,\phi,\gamma),
\end{equation}
where $E_d$ is photon energy in the restframe of disc coordinate $(r,\phi)$. From this, we calculate the observed specific disc flux by tracing rays from the disc to the observer following null-geodesics in the Kerr metric (using the code \textsc{ynogk}, which is based on \textsc{geokerr}: \citealt{Yang2013,Dexter2009}). 
A summary of the ray tracing procedure can be found in Appendix~\ref{app:model_raytracing} (or see e.g. \citealt{Ingram2019reltrans} for a more detailed description.)

\subsubsection{Illumination of the disc}
\label{sec:illum}

A precessing corona will preferentially illuminate different disc azimuths at different phases of its precession cycle. Instead of making assumptions about the shape of the corona, we follow \citet{Ingram2017h1743} by parameterising the QPO phase-dependent illuminating flux as a function of disc radius and azimuth with the emissivity function, $\epsilon(r,\phi,\gamma)$, such that
\begin{equation}
    \frac{I_\text{nt}(E_d,r,\phi,\gamma)}{D^2} = f_\text{R}(\gamma) N_\text{c}(\gamma) \epsilon(r,\phi,\gamma) \mathcal{R}(E_d),
    \label{eqn:Iref}
\end{equation}
where $D$ is the distance from the observer to the BH. Here, we normalise $\epsilon(r,\phi,\gamma)$ and $\mathcal{R}(E_d)$ (Eqs.~\ref{eqn:epsnorm} and \ref{eqn:Rnorm} in Appendix~\ref{app:model_norms}) such that $f_\text{R}(\gamma)$ is the \textit{observer's reflection fraction}. This is defined by \citet{Ingram2019reltrans} as the observed bolometric reflected flux divided by the directly observed bolometric coronal flux in the simplified case in which the disc re-emits the incident radiation isotropically. In this case, since $N_\text{c}(\gamma)$ is defined as the directly observed bolometric coronal flux, the observed bolometric reflected flux is simply $f_\text{R}(\gamma) N_\text{c}(\gamma)$. In reality, the reflected flux is \textit{not} emitted isotropically. The function $\mathcal{R}(E_d)$, which we discuss below, includes this subtlety.

We employ the following form for the emissivity function
\begin{equation}
\begin{split}
    \epsilon(r,\phi,\gamma) = \mathcal{N}_\epsilon \epsilon(r) \bigg\{ 1 + A_1 \cos^2\left[\frac{1}{2}(\gamma-\phi+\phi_1)\right] \\
    + A_2 \cos^2[\gamma-\phi+\phi_2] \bigg\},
    \label{eqn:eps1}
    \end{split}
\end{equation}
where the radial dependence is given by a twice broken power-law \citep{Wilkins2011, Wilkins2012}
\begin{equation}
\epsilon(r) =
\begin{cases} (r/r_\text{br,1})^{-q_1} &\mbox{if } r \le r_\text{br,1} \\ 
(r/r_\text{br,1})^{-q_2} & \mbox{if } r_\text{br,1} < r \le r_\text{br,2} \\
(r_\text{br,2}/r_\text{br,1})^{-q_2} (r/r_\text{br,2})^{-3} & \mbox{if } r > r_\text{br,2}.
\end{cases}
\label{eqn:eps2}
\end{equation}
This way, if $A_1=A_2=0$, the reflection spectrum will not depend at all on QPO phase, since the illumination pattern on the disc becomes axi-symmetric. When the asymmetric illumination (`asymmetry') parameters $A_1$ and $A_2$ are non-zero, there are instead bright patches on the disc that rotate around the disc rotation axis once per QPO cycle, with the location on the disc of the peak brightness set by the phase parameters $\phi_1$ and $\phi_2$. Specifically, $A_1 > 0$ and $A_2=0$ will lead to one bright patch rotating about the disc surface normal, and $A_1=0$ and $A_2 > 0$ will lead to two identical bright patches (see \citealt{Ingram2017h1743} for more details). The normalisation constant, $\mathcal{N}_\epsilon$, is set by Eq.~\ref{eqn:epsnorm}. We also parameterise the reflection fraction as a sum of sinusoids in the form of Eq.~\ref{eqn:Nc}.

\subsubsection{Thermal flux}
\label{sec:thermal}

We assume that the blackbody component of the disc specific intensity is given by
\begin{equation}
\frac{ I_\text{bb}(E_d,r,\phi,\gamma) }{ D^2 } = N_\text{d} ~B(E_d,T(r,\phi,\gamma)),
\end{equation}
where $N_\text{d}$ is a constant model parameter, $B(E,T)$ is the Planck function, and
\begin{equation}
    T(r,\phi,\gamma) = \bigg[ T_\text{visc}^4(r) + T_\text{irr}^4(r,\phi,\gamma) \bigg]^{1/4}.
    \label{eqn:T}
\end{equation}
Here, $T_{\rm visc}(r)$ is the `intrinsic' disc temperature; i.e. the temperature in the absence of irradiation. $T_{\rm irr}(r,\phi,\gamma)$ is the `irradiation' disc temperature; i.e. the temperature in the absence of viscous dissipation.

We set the intrinsic disc temperature as
\begin{equation}
T_\text{visc}(r) \propto \frac{ T_\text{visc,max} } { ( r^2 dA/dr )^{1/4} },
\end{equation}
where the constant of proportionality is set to ensure that the maximum temperature is $T_\text{visc,max}$ and the relativistic expression for $dA/dr$ is given by e.g. Equation A1 in \citet{Ingram2019reltrans}. In Newtonian gravity, $dA/dr=2\pi r$, and so the familiar $T_\text{visc} \propto r^{-3/4}$ emissivity of a simple disc model is recovered\footnote{Note that we do not employ a stress free inner boundary condition, which is appropriate if the corona is located inside of the disc, providing a torque.}. Note that any colour-temperature correction factor accounting for the spectrum not being strictly blackbody is simply swallowed up into the definition of $T_\text{visc,max}$.

The irradiation temperature is related to the thermalised portion of the illuminating flux via the Stefan-Boltzmann law. We parameterise it as
\begin{equation}
    T^4_\text{irr}(r,\phi,\gamma) \propto f_\text{R}(\gamma-\Delta \gamma) N_\text{c}(\gamma-\Delta \gamma) \epsilon(r,\phi,\gamma-\Delta\gamma).
      \label{eqn:kTirr_modulation}
\end{equation}
Defining $T_\text{irr,max}(\gamma)$ as the maximum value of $T_\text{irr}(r,\phi,\gamma)$ for a given QPO phase, we set the constant of proportionality in the above equation to ensure that the QPO phase-averaged value of $T_\text{irr,max}(\gamma)$ is proportional to the model parameter $kT_\text{i}$. Note that $kT_\text{i}$ effectively sets the fraction of the illuminating flux that thermalises in the disc atmosphere. The model parameter $\Delta\gamma$ accounts for the \textit{thermalisation timescale}, which is the time it takes for the irradiating flux to thermalise in the disc. This timescale is currently poorly understood, but it should lead to the thermal component responding to changes in the illuminating flux with a delay compared to the non-thermal component \citep[e.g. emission lines:][]{Garcia2013a}. Here we parameterise this thermalisation timescale such that the current disc temperature depends on what the irradiating flux was some time $\Delta\gamma /( 2\pi \nu_\text{qpo} )$ ago. This is an extremely simplified formalism. In reality, we may expect $T_\text{irr}(\gamma)$ to be smeared as well as delayed, and we may also expect the delay itself to depend on e.g. disc radius. 

Finally, we set the seed photon temperature in \textsc{nthcomp} to
\begin{equation}
    T_\text{bb}(\gamma) = \left[ T^4_\text{visc,max} + T^4_\text{irr,max}(\gamma) \right]^{1/4}.
    \label{eqn:seed_photon}
\end{equation}

\subsubsection{Non-thermal flux}
\label{sec:nonthermal}

We use the model \textsc{xillverCp} to calculate the non-thermal component of the restframe emergent disc spectrum, $\mathcal{R}(E_d)$. 
\textsc{xillverCp} calculates the emergent spectrum from a passive ($T_{\rm visc}=0$), constant density slab (electron number density $n_e=10^{15}~{\rm cm}^{-3}$) being irradiated by an \textsc{nthcomp} spectrum\footnote{Note that \textsc{xillverCp} is calculated for an irradiating spectrum given by \textsc{nthcomp} with the seed photon temperature hardwired to $kT_{bb}=0.05$ keV, whereas we allow our continuum spectrum to have a seed photon temperature that is free to vary with QPO phase.}. The output spectrum includes emission lines (most prominently the iron K$\alpha$ line), absorption edges (most prominently the iron K edge) and the Compton hump. It also includes a quasi-thermal component caused by some fraction of the irradiating photons thermalising in the disc, which we must ignore because we have already accounted for the thermalised illuminating flux in our blackbody component (see previous section). For \textsc{xillverCp}, this component peaks in the UV and is entirely below our bandpass, and so is simple to ignore.

An important input parameter of \textsc{xillverCp} is the ionisation parameter $\xi= 4\pi F_x/n_e$, where $F_x$ is the illuminating X-ray flux. We set the ionisation parameter from our existing parameterisation of the illuminating flux such that\footnote{Computing the ionisation in this way does mean the restframe reflection spectrum is non-uniform across the disc, thus is formally $\mathcal{R}(E_d,r,\phi,\gamma)$.}
\begin{equation}
    \xi(r,\phi,\gamma) \propto f_\text{R}(\gamma) N_\text{c}(\gamma) \epsilon(r,\phi,\gamma) / n_e(r).
    \label{xi_modulation}
\end{equation}
Defining $\xi_\text{max}(\gamma)$ as the maximum value of $\xi(r,\phi,\gamma)$ for a given QPO phase, we set the constant of proportionality in the above equation to ensure that the QPO phase-averaged value of $\xi_\text{max}(\gamma)$ is equal to the model parameter $\xi_0$. For $n_e(r)$ we adopt the form corresponding to Zone A of the \citet{Shakura1973} disc model, following e.g. \citet{Ingram2019reltrans,Mastroserio2019,Shreeram2020}: $n_e(r) \propto r^{3/2} [ 1 - \sqrt{r_\text{in}/r}]^{-2}$.

\chai{
\subsubsection{Approximations}

Our treatment of the restframe spectrum emergent from a given disc patch is very approximate. First of all, the \textsc{xillverCp} model that we use has $n_e = 10^{15} \text{cm}^{-3}$ hardwired, whereas we would theoretically expect the density of the disc in GRS~1915+105 to be closer to $n_e \sim 10^{20} \text{cm}^{-3}$ \citep{Shakura1973}. Second, the emergent spectrum is not strictly the sum of a thermal and a non-thermal component. In reality, the disc atmosphere is being irradiated from below by the intrinsic disc emission and from above by the corona, and the true emergent spectrum would need to be calculated by solving the radiative transfer equation with these boundary conditions \citep{Reis2008}. However, although high density \textsc{xillver} models are now available \citep{Garcia2016,Mastroserio2021}, they only consider irradiation by the corona and ignore the intrinsic disc emission. Moreover, the \textsc{xillver} solutions are calculated in steady-state, and so there is no thermalisation timescale. Our treatment is therefore the best time-dependent approximation of a high density irradiated disc with strong intrinsic emission currently available.
}

\subsection{The complete model}

The total observed QPO phase-dependent specific flux, $F(E,\gamma)$, is the sum of the disc and coronal contributions. 
\rmen{An example of the full model flux-energy spectrum for 8 QPO phases can be seen in Fig.~\ref{fig:nuprec_spectra}.  On top of the cut-off power law, the Compton hump can be seen at $\gtrsim10$~keV, along with the broadened iron K$\alpha$ line between $\sim5-\sim8$~keV.
Finally, the thermal flux can be seen at $\lesssim2$~keV.  This figure is computed from our results (as discussed in Section~\ref{sec:results}). The shape of the flux-energy spectrum varies with QPO phase because the asymmetry parameters, $A_1$ and $A_2$, are non-zero and other model parameters are modulated with QPO phase.}
Our model calculates $F(E,\gamma)$ for 16 QPO phases and Fourier transforms to get the QPO FT for the zeroth (phase-average), first, and second harmonics. 

Additionally, on top of the phase-dependent model described above, we include an extra \textsc{xillverCp} component to account for distant reflection. This component is fixed to be constant with QPO phase, as variations on the timescale of the QPO period should be strongly washed out by light-crossing delays for a distant reflector. Finally, we account for line-of-sight absorption with the model \textsc{tbabs}. The hydrogen absorption column $N_\text{H}$ is a free parameter, and we adopt the abundances of all other elements relative to hydrogen of \citet{Wilms2000}.  \rmen{These two components are not shown in Fig.~\ref{fig:nuprec_spectra}.}

\section{Model Fits}
\label{sec:fits}    

\subsection{Fitting procedure}

We fit our model simultaneously to 11 spectra overall\chen{.  As described in Section~\ref{sec:reconstructed_FT_spectra}, these are}: the flux-energy \textit{NICER}, FPMA and FPMB \textit{NuSTAR} spectra, plus the first and second harmonics of the real and imaginary parts of the QPO FT measured separately by \textit{NICER} and \textit{NuSTAR}. Using \textsc{xspec} version 12.10.1f, we applied the same \textit{NICER} response matrix to all 5 \textit{NICER} spectra, we used the FPMA response for the flux-energy FPMA spectrum, and the FPMB response for both the FPMB flux-energy spectrum and the \textit{NuSTAR} QPO FT. This is because the \textit{NuSTAR} QPO FT was calculated using FPMB channels as the subject bands, and the full band FPMA light curve as the reference band. We see from Fig.~\ref{fig:ratio} that the cross-calibration between \textit{NICER} and the \textit{NuSTAR} modules is good, except for the normalisation. We account for this by multiplying our model by a floating constant, which is fixed to unity for \textit{NICER} and left free for the \textit{NuSTAR} FPMA and FPMB. 

As we used different reference bands for the calculation of the \textit{NICER} and \textit{NuSTAR} QPO FTs (the full band \textit{NICER} and \textit{NuSTAR} FPMB light curves respectively), there is a phase difference between them caused by the phase lag between the two reference bands (see Section~\ref{sec:lags}). We account for this by setting a phase offset $\phi_\text{c}$ for the first harmonic (the phase offset for the second harmonic is exactly $2\phi_\text{c}$\footnote{This follows as the phase of each harmonic in every energy band is measured in reference to the phase of the first harmonic in the reference band.  $\phi_\text{c}$ is the phase offset between the first harmonics in the reference band between the two instruments, and the same offset is $2\phi_\text{c}$ when instead measured at frequency of the second harmonic.  See Appendix~\ref{app:phi_c} for the derivation.}). Similarly to floating calibration constants employed for flux-energy spectra, we set $\phi_\text{c}=0$ for \textit{NICER} and leave $\phi_\text{c}$ as a free parameter for \textsc{\textit{NuSTAR}}.

In our fits, we leave the truncation radius $r_\text{in}$ as a free parameter and fix the spin to $a=0.998$. While we do not necessarily expect that this choice reflects the true value of the spin \cite[see e.g.][]{mills2021}, this value enables the widest possible range of the $r_\text{in}$ parameter to be explored without it becoming smaller than the ISCO. Although the spin does affect the geodesics, this only has a very subtle effect on the spectrum compared with the location of the disc inner radius.

\chen{
The full list of the free parameters in our model is included in Table~\ref{tab:auto_temp_free_pars}, with those modulated with QPO phase $\gamma$ in the manner of Eq.~\ref{eqn:Nc} grouped together, and are labeled in the manner of `$N_\text{c}(\gamma)$'. We begin by finding a global best fit through minimising the $\chi^2$ fit statistic, before running a long Markov ChainMonte Carlo (MCMC) simulation around the best fit to explore the parameter space and understand the significance of the best fitting parameter values.  
}

\subsection{Results}
\label{sec:results}

\begin{table}
    \begin{tabular}{lllll}
\toprule
\multicolumn{2}{c}{ \chen{Parameter} }                                  &	Unit                        &	Chain mean             &	Description  \\                            
\midrule
                                       &	$N_\text{H}$           &	$10^{22}$                   &	$6.8\pm0.3$            &	\textsc{tbabs} col. density \\             
\cline{1-2}
\multirow{5}{*}{$\Gamma(\gamma)$}      &	$A_{1\Gamma}$          &	                            &	$0.04\pm0.01$          &	\multirow{5}{*}{Photon index} \\           
                                       &	$A_{2\Gamma}$          &	                            &	$0.05\pm0.02$          &	\\                                         
                                       &	$\phi_{1\Gamma}$       &	cyc                         &	$0.42\pm0.04$          &	\\                                         
                                       &	$\phi_{2\Gamma}$       &	cyc                         &	$0.13\pm0.03$          &	\\                                         
                                       &	$\Gamma$               &	                            &	$1.93\pm0.01$          &	\\                                         
\cline{1-2}
\multirow{5}{*}{$kT_\text{e}(\gamma)$} &	$A_{1e}$               &	keV                         &	$38. \pm13. $          &	\multirow{5}{*}{Electron temp.} \\         
                                       &	$A_{2e}$               &	keV                         &	$19. ^{+8. }_{-9. }$   &	\\                                         
                                       &	$\phi_{1e}$            &	cyc                         &	$0.44\pm0.04$          &	\\                                         
                                       &	$\phi_{2e}$            &	cyc                         &	$0.97\pm0.03$          &	\\                                         
                                       &	$kT_\text{e}$          &	keV                         &	$57. ^{+9. }_{-10. }$  &	\\                                         
\cline{1-2}
                                       &	$\log\xi^\dagger$      &	                            &	$2.5\pm0.2$            &	Ionisation of dist. refl.\\                
                                       &	norm$^\dagger$         &	$10^{-3}$                   &	$0.7\pm0.2$            &	Normalisation of dist. refl.\\             
                                       &	$\text{A}_\text{Fe}$   &	$\text{A}_{\text{Fe}\odot}$ &	$7. \pm1. $            &	Accreting Fe abundance\\                   
                                       &	Incl.                  &	deg                         &	$75.1^{+0.5}_{-0.3}$   &	Inclination of source\\                    
                                       &	$r_\text{in}$          &	$\text{r}_\text{g}$         &	$1.43^{+0.01}_{-0.02}$ &	Inner truncation radius\\                  
                                       &	$q_1$                  &	                            &	$13.6^{+0.8}_{-0.7}$   &	Inner emissivity index\\                   
                                       &	$r_\text{br,1}$        &	$r_\text{in}$               &	$2.4\pm0.5$            &	$1^\text{st}$ break radius\\               
                                       &	$q_2$                  &	                            &	$4. ^{+5. }_{-4. }$    &	Outer emissivity index\\                   
                                       &	$r_\text{br,2}$        &	$r_\text{br,1}$             &	$14. ^{+11. }_{-10. }$ &	$2^\text{nd}$ break radius\\               
\cline{1-2}
\multirow{4}{*}{Asym.}                 &	$A_1$                  &	                            &	$0.5\pm0.3$            &	\multirow{4}{*}{Asymmetric illumination}\\ 
                                       &	$A_2$                  &	                            &	$1.3\pm0.7$            &	\\                                         
                                       &	$\Phi_1$               &	cyc                         &	$0.9\pm0.1$            &	\\                                         
                                       &	$\Phi_2$               &	cyc                         &	$0.99^{+0.06}_{-0.05}$ &	\\                                         
\cline{1-2}
\multirow{5}{*}{$f_\text{R}(\gamma)$}  &	$A_{1f}$               &	                            &	$0.3\pm0.1$            &	\multirow{5}{*}{Reflection fraction}\\     
                                       &	$A_{2f}$               &	                            &	$0.3\pm0.1$            &	\\                                         
                                       &	$\phi_{1f}$            &	cyc                         &	$0.45\pm0.05$          &	\\                                         
                                       &	$\phi_{2f}$            &	cyc                         &	$0.94\pm0.05$          &	\\                                         
                                       &	$f_\text{R}$           &	                            &	$0.96^{+0.08}_{-0.09}$ &	\\                                         
\cline{1-2}
                                       &	$\log\xi_\text{max}$   &	                            &	$4.26^{+0.08}_{-0.09}$ &	Max radial ionisation\\                    
                                       &	$\Delta\gamma$         &	cyc                         &	$0.17\pm0.07$          &	Thermalisation phase lag\\                 
                                       &	$kT_\text{v,max}$      &	keV                         &	$0.24\pm0.05$          &	Max radial viscous heating\\               
                                       &	$kT_\text{i}$          &	keV                         &	$0.64\pm0.04$          &	Heating from irradiation\\                 
                                       &	$N_\text{d}$           &	                            &	$25. \pm11. $          &	Disc normalisation\\                       
\cline{1-2}
\multirow{5}{*}{$N_\text{c}(\gamma)$}  &	$A_{1N}$               &	                            &	$1.2\pm0.2$            &	\multirow{5}{*}{Coronal Normalisation}\\   
                                       &	$A_{2N}$               &	                            &	$0.8\pm0.3$            &	\\                                         
                                       &	$\phi_{1N}$            &	cyc                         &	$0.96\pm0.03$          &	\\                                         
                                       &	$\phi_{2N}$            &	cyc                         &	$0.70\pm0.03$          &	\\                                         
                                       &	$N_\text{c}$           &	                            &	$5.8\pm0.2$            &	\\                                         
\cline{1-2}
                                       &	FMPA                   &	                            &	$0.944\pm0.001$        &	\textit{NuSTAR} FMPA norm.\\               
                                       &	FMPB                   &	                            &	$0.951\pm0.001$        &	\textit{NuSTAR} FMPB norm.\\               
                                       &	$\phi_\text{c}$        &	cyc                         &	$0.018\pm0.004$        &	\textit{NuSTAR} phase offset\\             
\bottomrule
\end{tabular}

    \centering
    \caption{The mean and $\pm1\sigma$ credible interval of the posterior distributions of each parameter, calculated with a MCMC.
     The two parameters marked with $^\dagger$ are those solely relating to the phase-constant reflected component assumed to come from a distant reflector.  The `Asym.' parameters are those that govern the asymmetric illumination profile. The FPMA and FPMB norm parameters are floating calibration constants.}
    \label{tab:auto_temp_free_pars}
\end{table}

\begin{figure*}
    \includegraphics[width=\textwidth]{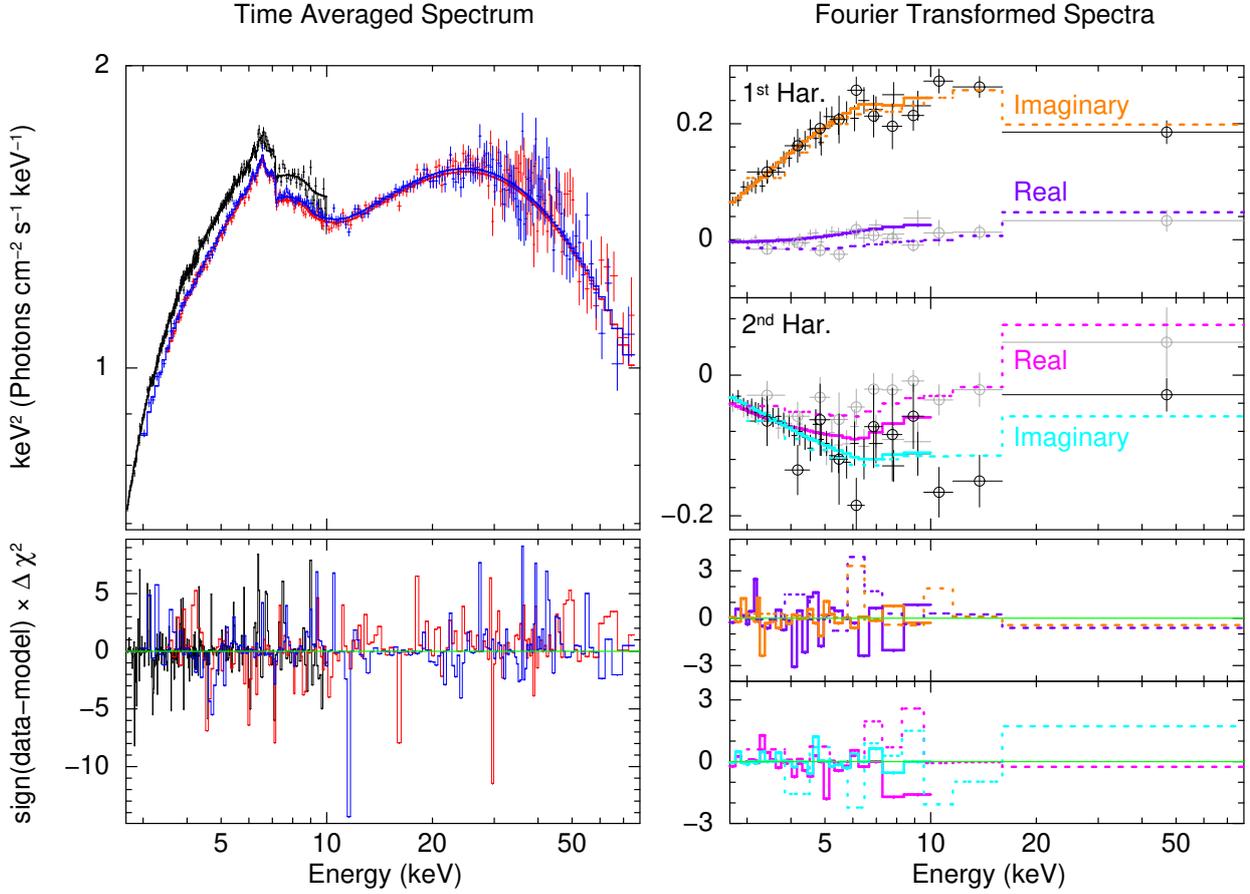}
    \centering
    \caption{The unfolded best fit model, and the ratio of the data to the model, for the flux-energy spectra (left) and  the harmonics of the QPO (right). The flux-energy spectra from \textit{NICER} is in black, \textit{NuSTAR} FPMA is in red, and \textit{NuSTAR} FPMB is in blue, which have been rebinned for plotting purposes. 
    The spectra of the real and imaginary components of the $1^\text{st}$ QPO harmonic are in \chen{purple} and orange respectively.  Likewise, the $2^\text{nd}$ QPO harmonic components are shown in magenta and light blue. Both harmonics have the \textit{NuSTAR} data points with circles and a dotted line, with the circle-less points and solid lines corresponding to \textit{NICER}.  \chen{The markers in grey show the real component of the FT spectra, whereas those in black show the imaginary component.} The axis labels apply to both the left and right hand sides of the figure, including the units\chen{; this} follows from the Fourier transformed spectra being the RMS multiplied by a unitless phase term.}
    \label{fig:auto_temp_model_fits}
\end{figure*}

\begin{figure}
	\centering
	\includegraphics[]{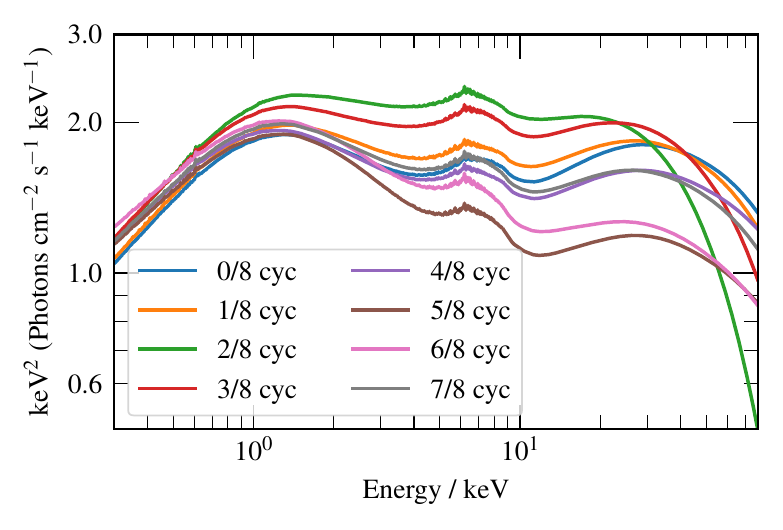}
    \caption{The QPO phase dependent component of our best-fitting model; without line-of-sight absorption or the distant reflector component.  This is comprised of the \textsc{nthcomp} cut-off powerlaw; a ray-traced \textsc{xillverCp} reprocessed spectrum with relativistically smeared atomic features ($~5-8$~keV) and Compton hump ($\gtrsim20$~keV); plus a blackbody thermal spectrum at $\lesssim2$~keV.  This shows the model at 8 phases, although 16 phases are used for the calculation of the Fourier transformed components.}
    \label{fig:nuprec_spectra}
\end{figure}

Our global best fit, which is shown in Fig.~\ref{fig:auto_temp_model_fits}, has $\chi^2=3374.6$ for 3179 degrees of freedom (DoF). 
The phase-average flux-energy spectrum in the upper-left of the figure shows the good fit, especially to the Fe~K$\alpha$ line. The right-hand side of the figure shows the real and imaginary parts of the FT spectra of the first two harmonics. The features of these spectra are markedly less intuitive, and so we rely on analysis of the parameter space to understand the model fit.  As the distant reflector is constant on the QPO timescale, the FT spectra do not
include this component.  However, as they are normalised to the observed flux (see Eq.~\ref{eqn:FT_spectra}), the absorption column does matter and therefore is included.

We explored parameter space by running a long \chen{MCMC} simulation with 120 walkers which each run for 250,000 steps from an initial distribution based on the covariance matrix of the best fit.  Every parameter had a uniform prior with bounds considerably far from the range required, except where parameters must be non-negative\footnote{The amplitude `$A$' parameters of phase-modulated quantities are such examples, including the asymmetry parameters $A_1$ and $A_2$.}. We burn the first 50,000 steps of each walker, enough to ensure that the Geweke convergence diagnostic \citep{Geweke1992} is within $\pm0.3$ for every parameter and we therefore only consider steps where the MCMC has converged. Finally, we thin the MCMC down, only taking every $100^\text{th}$ step from each walker.
We show the posterior means and $1\sigma$ credible intervals from the MCMC in Table~\ref{tab:auto_temp_free_pars}.

\begin{figure}
	\centering
	\includegraphics[]{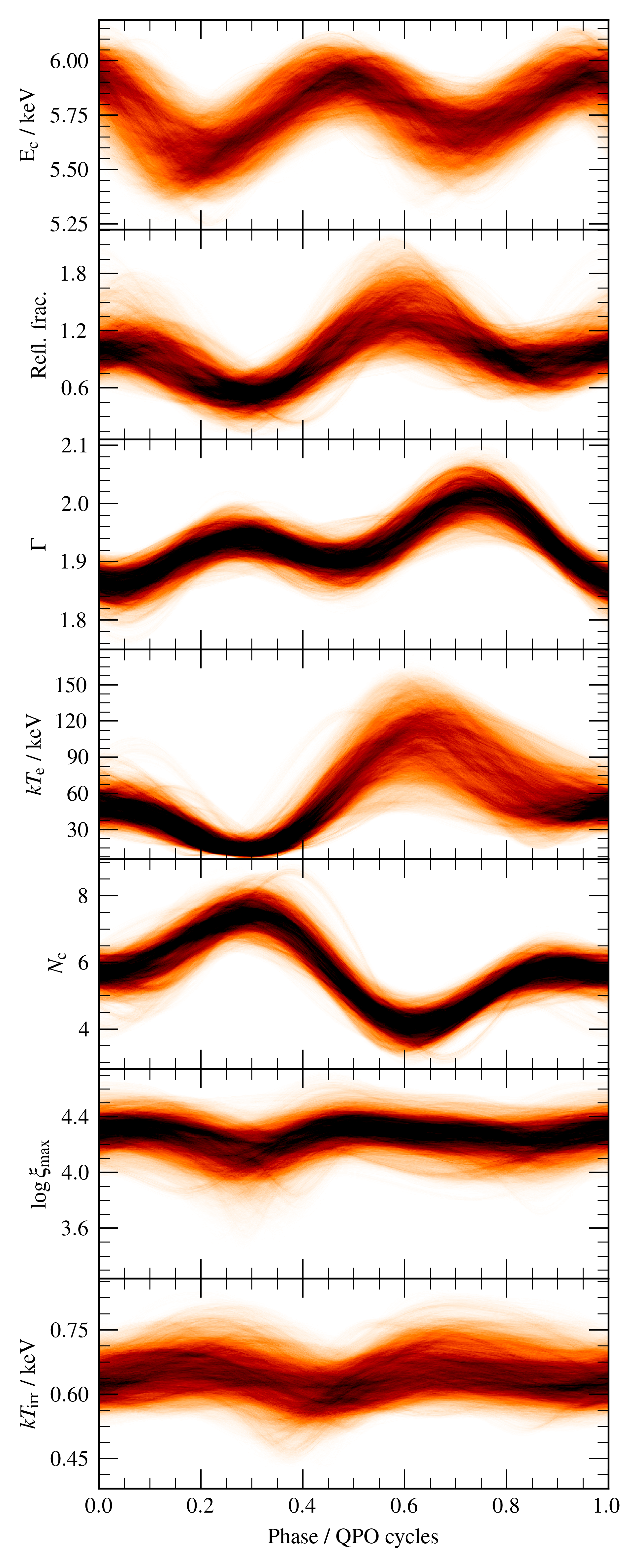}
    \caption{Curves produced from the parameters from steps in the MCMC.  The iron-line centroid energy $E_\text{c}$ was calculated from the varying illumination profile, assuming a rest-frame $\delta-$function profile (see text for details).  The reflection fraction, $\Gamma$, $kT_\text{e}$, and $N_\text{c}$ are the modulations straight from their parameters in the model, whereas $\log\xi_\text{max}$ and $kT_\text{irr}$ are calculated from the irradiating flux, as described in the text.}
    \label{fig:auto_temp_hists}
\end{figure}

In Fig.~\ref{fig:auto_temp_hists}, we visualise our results by reconstructing parameter modulations from the thinned chain. For each step in the chain (there are $240,000$ steps altogether), we use the parameter values corresponding to that step in order to calculate each of the 7 quantities plotted in Fig.~\ref{fig:auto_temp_hists} as a function of QPO phase. From these $240,000$ functions of QPO phase, we create a 2D histogram, which we plot as a probability map (black represents the largest probability). 

Panels 2-5 in Fig.~\ref{fig:auto_temp_hists} show the parameters that are allowed to vary with QPO phase via a sum of sinusoids (e.g. Eq.~\ref{eqn:Nc}); from top to bottom: $f_\text{R}(\gamma)$, $\Gamma(\gamma)$, $kT_\text{e}(\gamma)$ and $N_\text{c}(\gamma)$. 
We consider the posterior distributions in Fig.~\ref{fig:param_modulation_corner_plots}, and we see that all the parameter modulations are significant to at least $3\sigma$, with $N_\text{c}(\gamma)$ likely at a much higher significance.
The bottom two panels in the figure are $\log\xi_\text{max}(\gamma)$ and $kT_\text{irr}(\gamma)$. The modulations in these parameters are not free to vary in our fit, they are instead calculated from $N_\text{c}(\gamma)$ and $f_\text{R}(\gamma)$ (see Eqs.~\ref{eqn:kTirr_modulation}, and \ref{xi_modulation}).  These are remarkably constant with QPO phase, as they are both $\propto N_\text{c}(\gamma)f_\text{R}(\gamma)$ (neglecting the phase shift $\Delta\gamma$), whose modulations are approximately out of phase.

\begin{figure}
    \centering
    \includegraphics{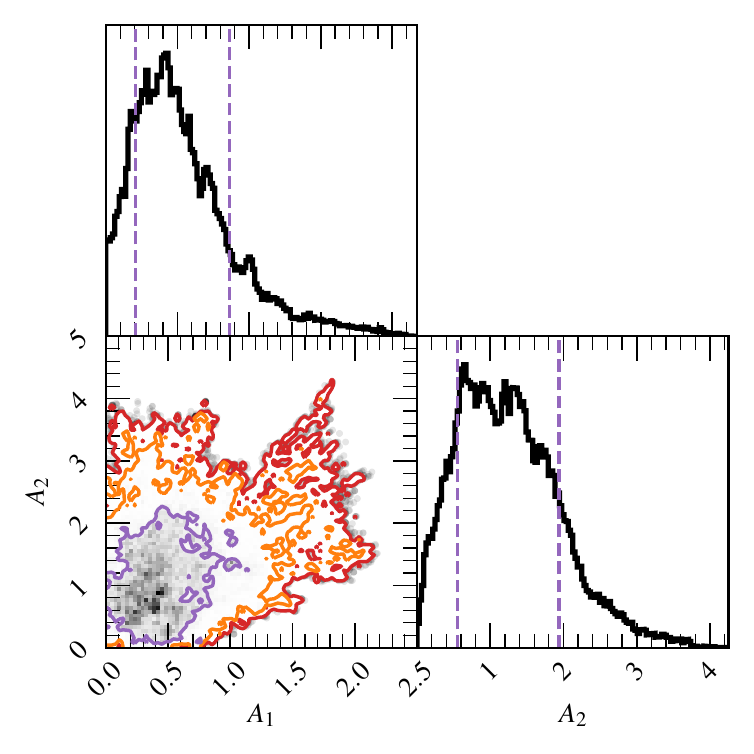}
    \caption{$A_1$ and $A_2$ contour based around the MCMC.  The 1, 2, and 3 $\sigma$ credible intervals are shown as contours (purple, orange, red), with the blue lines highlighting the values at the best-fit. Within the $1\sigma$ contour the density is shown as a gray-scale 2D-histogram. Outside the $3\sigma$ contour individual points are shown as grey points. 
    The marginalised histograms also show the $\pm1\sigma$ credible interval with purple dashed lines.}
    \label{fig:auto_temp_a1_a2}
\end{figure}

The top panel is iron line centroid energy. In order to determine this, we first calculate the observed QPO phase-dependent reflection spectrum (Eq.~\ref{eqn:Fd}) assuming a $\delta-$function iron line in the disc restframe (i.e. $I(E_\text{d})=\delta(E_\text{d}-6.4\text{keV})$) and then calculate the centroid energy of the resulting QPO phase-dependent line profile (using Equation 7 from \citealt{Ingram2017h1743}). Any modulations of this function with QPO phase are caused entirely by QPO phase dependence of the emissivity function $\epsilon(r,\phi,\gamma)$, which in turn is driven exclusively by the asymmetry parameters $A_1$ and $A_2$. We see that the line centroid energy is strongly modulated with QPO phase, implying that the emissivity function is required by the fit to vary with QPO phase. 
We show the posterior distribution of the asymmetry parameters from the MCMC in Fig.~\ref{fig:auto_temp_a1_a2}.  The point $A_1=A_2=0$ lies outside of the $2\sigma$ contour (in red), therefore we are able to reject the axi-symmetric null-hypothesis of $A_1=A_2=0$ at the $2\sigma$ significance level.

We show the QPO phase-dependent spectrum of the best-fitting model in Fig.~\ref{fig:nuprec_spectra}, without the line-of-sight absorption and distant reflector. We see large changes in spectral shape over the course of the 8 QPO phases pictured, most obviously the change in continuum normalisation $N_\text{c}(\gamma)$ following the same trend as shown in Fig.~\ref{fig:auto_temp_hists}.  As the normalisation of the reflection spectrum is broadly constant with phase, its key features of the Fe~K$\alpha$ and Compton hump are less pronounced when $N_\text{c}(\gamma)$ is larger.  The temperature of the Compton hump itself does change, most noticeably at its lowest value of $\gamma\approx1/4$~cycles.

When we consider the covariance between the modulated variables (see Fig.~\ref{fig:param_harmonics_corner_plots}), we see the strongest correlation is between the reflection fraction $f_\text{R}$ and continuum normalisation $N_\text{c}$.  Interestingly, the phase-averaged values are negatively correlated, however the size and phases of the modulations are positively correlated.  We can see in Fig.~\ref{fig:auto_temp_hists} that these modulations are also in anti-phase, and so their modulations work against each other to keep the incident flux onto the disc approximately constant,
as can be seen in the waveforms of $\log\xi_\text{max}(\gamma)$ and $kT_\text{irr}(\gamma)$.

\begin{figure}
    \centering
    \includegraphics{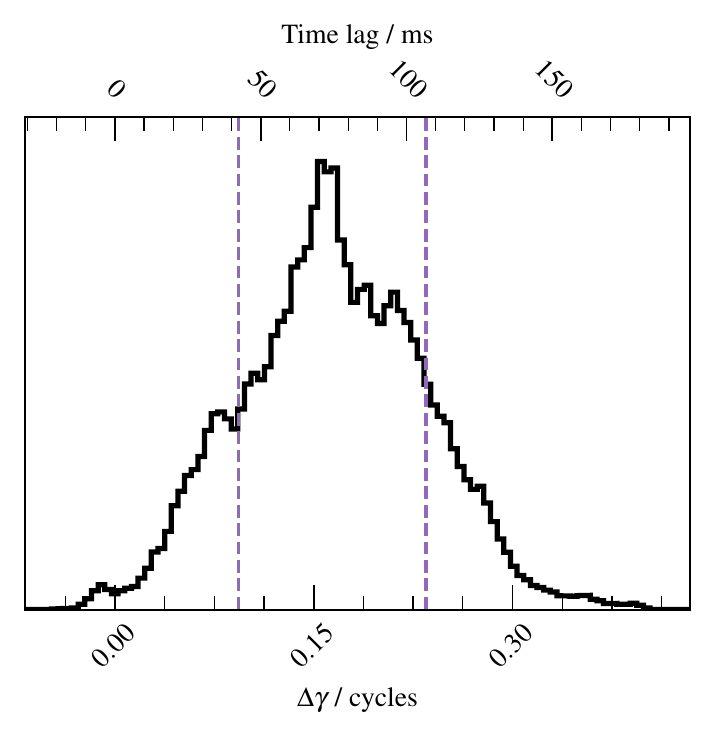}
    \caption{Histogram of $\Delta\gamma$ from the MCMC, with the corresponding time-lag for a frequency of 2.2~Hz.  The $\pm1\sigma$ credible interval is outlined with purple dashed lines.}
    \label{fig:auto_temp_delgam}
\end{figure}

We find that the posterior mean of the phase lag $\Delta\gamma$, intended to represent the time it takes photons to thermalise in the disc atmosphere, is $\Delta\gamma \approx 0.17$ QPO cycles. Fig.~\ref{fig:auto_temp_delgam} shows the posterior distribution of $\Delta\gamma$ from the MCMC, including a conversion from QPO cycles to time lag for a QPO frequency of $\nu_\text{QPO}=2.2$~Hz. 
This phase lag can be seen in the bottom two panels of Fig.~\ref{fig:auto_temp_hists}, since the dip in $kT_\text{irr}(\gamma)$ occurs $\sim 0.17$ QPO cycles after the dip in $\log\xi_\text{max}(\gamma)$. It can also be seen in Fig.~\ref{fig:nuprec_spectra}: e.g. the orange line for QPO phase $=0.125$ cycles corresponds to a peak in iron line centroid energy but not to a peak in disc peak temperature.
Converting $\Delta\gamma$ to a time lag gives $\sim 75$~ms, which is rather large. This large value is possibly due to model systematics, since if the thermalisation timescale really were this long, then we would see much longer thermal reverberation lags than have been observed \citep[e.g.][]{Uttley2011, kara2019}.

\section{Discussion}
\label{sec:discussion}

We have conducted a phase-resolved spectral analysis of a $\sim 2.2$ Hz QPO from GRS~1915+105 observed simultaneously by \textit{NICER} and \textit{NuSTAR}. 
We found that the continuum normalisation $N_\text{c}(\gamma)$, photon index $\Gamma(\gamma)$, reflection fraction $f_\text{R}(\gamma)$, and electron temperature $kT_\text{e}(\gamma)$ are all required to be modulated with QPO phase to $>3\sigma$ confidence, plus we rule out that the asymmetric illumination parameters $A_1=A_2=0$ at the $2\sigma$ confidence level.  Alongside this, we found a small inner truncation radius $r_\text{in}$, and a large thermalisation phase lag $\Delta\gamma$.
We now discuss the implications of these results.

\subsection{Comparison with H1743-322}

We compare our results with the previous similar study of a $\sim0.25$~Hz type-C QPO in H~1713-322 by \citet{Ingram2017h1743} (hereafter I17).
The first thing to note is the strikingly similar profiles of the continuum normalisation, $N_\text{c}(\gamma)$ (compare our Fig.~\ref{fig:auto_temp_hists} with the Fig.~5 in I17), which is a proxy for the X-ray flux. Whereas the fractional amplitude of the $N_\text{c}(\gamma)$ modulation is larger here, the profiles -- and therefore the QPO waveforms -- are similar for the two observations, despite the QPO frequencies being very different. This similarity is confirmed by the measured phase difference between harmonics, $\psi$, which defines the QPO waveform \citep{Ingram2015grs1915}. Here we measure $\psi/\pi \approx 0.2$ for the \textit{NuSTAR} observation, whereas the \textit{NuSTAR} measurement for H~1743-332 was $\psi/\pi \approx 0.3$ \citep{Ingram2016}. This indicates that the $N_\text{c}(\gamma)$ modulation for H~1743-322 should be similar to what we measure here for GRS~1915+105 but with the peak slightly delayed, which can indeed be seen in the $N_\text{c}(\gamma)$ profiles. It is perhaps surprising that the QPO waveform of the two observations is so similar given that \citet{Ingram2015grs1915} observed it to change dramatically between two \textit{RXTE} observations of GRS~1915+105 with QPO frequencies of $\sim 0.5$ Hz and $\sim 2.25$ Hz. Fig. 5 (right panel) in \cite{deRuiter2019} reveals that this is because $\psi$ reduces more steeply with QPO frequency in GRS~1915+105 than it does in H~1743-322, such that its value for GRS~1915+105 at $\nu_\text{qpo} \sim 2$ Hz is close to the corresponding value for H~1743-322 at $\nu_\text{qpo} \sim 0.2$ Hz. The measurement of $\psi$ presented here is therefore consistent with the trends found by \citet{deRuiter2019}. Beyond the overall shape, we also see a much wider spread in our $N_\text{c}(\gamma)$ histogram than that presented by I17. This is because we are now using a much more flexible model than in previous work, including modulations in $kT_\text{e}$, and so it is better able to compensate when $N_\text{c}(\gamma)$ deviates from its best fitting functional form.

As for the modulation in iron line centroid energy, $E_\text{c}(\gamma)$, we see a strong second harmonic (evidenced by there being two maxima per QPO phase as opposed to one) both in our observation and in the I17 analysis of H~1743-322. This property was also previously observed for GRS~1915+105 by \citep{Ingram2015grs1915}, albeit with a low statistical significance. 
We however see that the phase of the $E_\text{c}(\gamma)$ modulation is shifted here with respect to the H~1743-322 observation: here the maxima in $E_\text{c}$ occur at QPO phase $\gamma \sim 0$ and $0.5$ cycles, whereas for H~1743-322 they occur at $\gamma \sim 0.2$ and $\sim 0.7$ cycles. A similar evolution in the $E_\text{c}(\gamma)$ waveform is seen between the two GRS~1915+105 observations presented in \citet{Ingram2015grs1915}: their $E_\text{c}(\gamma)$ waveforms for $\nu_\text{qpo} \sim 0.5$ Hz and $\nu_\text{qpo} \sim 2.25$ Hz QPOs are respectively similar to the $\sim 0.2$ Hz QPO in H~1743-322 and the $\sim 2.2$ Hz QPO in GRS~1915+105 presented here. This gives a potential hint that the phase of the line centroid energy modulation evolves systematically with QPO frequency, as is the case for the flux modulation \citep{deRuiter2019}. We additionally see that the $E_\text{c}(\gamma)$ modulation presented here has a lower mean and a higher amplitude than that presented by I17. This is consistent with the disc inner radius reducing as the QPO frequency increases: the mean is reduced by increased gravitational redshift and the amplitude is increased by faster orbital motion closer to the BH.

I17 found modulations of the reflection fraction, $f_\text{R}$, and power law index, $\Gamma$, to have $3.52 \sigma$ and $0.95 \sigma$ significance respectively, 
whereas here we find both modulations to have $>3\sigma$ significance.
As the reflected spectrum is spectrally harder than the directly observed spectrum, an increase in spectral hardness can be caused either by a reduction in $\Gamma$ or an increase in $f_\text{R}$.  Whereas the $\Gamma(\gamma)$ and $f_\text{R}(\gamma)$ modulations presented by I17 are broadly in phase, suggesting they somewhat compensate for each other. Here we find they are broadly in anti-phase suggesting they are both contributing to the modulation of the spectral hardness.

\subsection{Asymmetric illumination profile}

Our model requires an asymmetric illumination profile with $A_1=0.5\pm0.3$ and $A_2=1.3\pm0.7$ which is consistent with two, nonidentical bright patches rotating around the disc with QPO phase.
The QPO phase dependence of the iron line profile could, to some extent, be reproduced by changes in the ionisation state of the disc atmosphere, which would cause changes in the shape and centroid energy of the iron line in the restframe reflection spectrum (since e.g. Compton broadening of the line increases with the number of free electrons and higher order ions are more tightly bound and therefore produce higher energy fluorescence lines). The parameters that affect the shape of the restframe reflection spectrum and are modulated with QPO phase in our model are $\Gamma$, $kT_\text{e}$, and $\log\xi$. All three affect the ionisation state of the disc, with larger $kT_\text{e}$, larger $\log\xi$, and smaller $\Gamma$ increasing the number of irradiating photons with a high enough energy to ionize neutral iron ($E > 7.1$ keV). In the model employed by I17, only $\Gamma$ was allowed to vary with QPO phase, whereas here we also allow $kT_\text{e}$ and $\log\xi$ to vary, with $\log\xi$ tied to the illuminating flux. 
We also note that here we employ a radial $\log\xi$ profile, whereas only a single ionisation parameter was used for the entire disc in I17.
Given the flexibility of our model and the conservative nature of our analysis, we find that the asymmetric illumination parameters $A_1$ and $A_2$ are only required to be non-zero at $>2\sigma$ confidence level.

\chai{Such an asymmetric, QPO phase-dependent illumination pattern is naturally expected if the corona illuminating the disc is precessing \citep{Ingram2012}. Alternatively, precession of the disc itself (i.e. precession of the reflector and not the illuminator) could potentially explain our data \citep{schnittman2006}. However, disc precession }
is not expected theoretically \citep{bardeen1975,liska2019} unless the frame dragging effect is strong enough to tear the disc into a number of discrete, independently precessing rings \citep{Nixon2012, Liska2020}. Such a configuration could in principle cause QPOs if the number of independent rings is small enough to produce a coherent oscillation. However, disc tearing requires a large misalignment between the binary and BH rotation axes, which is expected to be rare since the biggest natal kicks, which would produce the largest mis-alignments, are also the most likely to completely disrupt the binary system \citep{fragos2010}. This is in contrast to the near-ubiquity of Type-C QPOs. Line profile modulations
could \chai{also} result from c-mode disco-seismic waves \citep[e.g.][]{kato1980, kato2001}. However, this would not explain the QPO being much stronger in the Comptonised spectrum than in the disc spectrum, and the line profile variations produced by c-modes are quite subtle compared with what we observe \citep[see e.g. Fig 5 in][]{tsang2013}, due to the c-mode oscillation only occuring in a narrow range of disc radii. The spiral waves of the AEI model are \chai{also} consistent with the asymmetric illumination profile we measure here \citep{Varniere2002}.
However, the AEI model is not consistent with a reflection fraction modulation, \chai{which we measure with} $>3\sigma$ \chai{significance}, whereas the precessing corona model is.

A new QPO diagnostic will soon be available in the form of X-ray polarization. Whereas the precession model predicts the polarization degree and angle to be modulated on the QPO period \citep{Ingram2015polarization}, alternative models such as the AEI do not. After its launch later this year, the \textit{Imaging X-ray Polarimetry Explorer} (\textit{IXPE}) should be capable of detecting QPOs in the X-ray polarization properties in a long exposure observation of a bright X-ray binary, as long as high inclination X-ray binaries turn out to be moderately polarized \citep[$\gtrsim4\%$;][]{ingram2017observational}.

\subsection{Inner Radius}

Our best-fitting model also requires a small disc inner radius of $r_\text{in}\simeq1.43^{+0.01}_{-0.02}~\rg$, where $\rg=GM/c^2$.  While this is compatible with previous reflection spectroscopy results \citep{Miller2013,Zhang2019,Shreeram2020}, it does not leave much room for a corona to precess within this truncation radius in order to give rise to the best fitting asymmetric illumination profile. \chai{Moreover, it is very much incompatible with the precession being specifically at the Lense-Thirring precession frequency, which requires $r_{\rm in} \approx 41~\rg$ to reproduce a QPO frequency of $\approx 2.2$ Hz\footnote{\chen{Using Eq.~2 of \cite{Ingram2009} with $\zeta=0$, as seen in numerical simulations \citep{fragile2007}, $r_{\rm in}=r_{\rm isco}(a)$, $a=0.998$, and $M=10.1~\text{M}_\odot$ \citep{Steeghs2013}.}}.

}

\chai{We therefore first consider if our analysis has returned an artificially small disc inner radius due to the use of an inadequate continuum model.}
Including a second, softer power law component can yield a larger truncation radius in fits to the flux-energy spectrum, since the new component takes the place of the broad red wing characteristic of small $r_\text{in}$ \citep{Yamada2013,Mahmoud2018,Mahmoud2019,Zdziarski2021accretion, Zdziarski2021does}.
Such a treatment could therefore yield a larger truncation radius in our analysis if our measured disc inner radius is driven primarily by the time-averaged iron line profile. However, our analysis is also sensitive to the QPO phase-dependence of the iron line profile.
For a large truncation radius, the red wing can dominate over the blue wing when the receding disc material is illuminated, whereas for a small truncation radius the variability is entirely in the blue wing because the red wing is always suppressed by gravitational redshift \citep{Ingram2012,You2020}.

By excluding the iron line from the flux-energy spectrum, but leaving in those bands for the Fourier transformed spectra, we are able to investigate whether this small $r_\text{in}$ is required by only the phase-average spectrum or also by the QPO phase dependence of the spectrum. We therefore run a new phase-resolved fit in which we ignore the $5-8$~keV data from the flux-energy spectrum but leave in those bands for the Fourier transformed spectra.
For this new fit, the measured $r_\text{in}$ value will not be driven by the time-averaged shape of the iron line, but by how its shape \textit{changes} throughout the course of each QPO cycle.
This new fit also returns a small disc inner radius of $r_\text{in}=1.49_{-0.04}^{+0.03}~\rg$. It is still possible that including a more complex continuum could allow for a slightly larger $r_\text{in}$, but this would add many more degrees of freedom to our already extremely flexible model.

\chai{Our fits therefore indicate (although not definitively) asymmetric illumination of a disc extending to a very small inner radius. It could be that the corona is 
not radially extended but instead} a vertically extended structure, such as a precessing jet \citep{Kalamkar2015,Stevens2016,Kylafis2020} (if the jet base is sufficiently X-ray bright; \citealt{Fender1999, markoff2005}). GRMHD simulations show that jet precession can be induced by the frame dragging effect \citep{liska2018}, but only so far in the presence of a thick disc. Indeed, our best fitting model includes a very strongly peaked emissivity function, with $q_1=13.6^{+0.8}_{-0.7}$ for disc radii $r<3.3\pm0.6~\rg$. 
This emissivity profile is roughly compatible with one created by a vertically extended corona raised slightly above the thin accretion disc \citep{Wilkins2012}.

\chai{It is, however, unclear exactly how the frame dragging effect could drive such slow precession for such a small disc inner radius. Perhaps the corona is vertically extended with the density increasing with distance from the BH. This weighting of the density to larger distances from the BH will slow down precession compared with a precessing ring at the same $r_\text{in}$. The torque exerted on the corona by the outer disc, which is not currently considered in calculations of the precession frequency, will also slow down precession. It is not, however, clear whether or not these two effects are sufficient to solve the discrepancy we find here. Even if they are, and Lense-Thirring precession really is the true type-C QPO mechanism, their presence will make it much more difficult to infer BH mass and spin from the QPO frequency than previously hoped. A clear counter-argument to this point is provided by the QPO triplet in GRO J1655-40, the frequencies of which can be explained very well by the relativistic precession model, returning a precise mass prediction that agrees with the dynamical value \citep{Motta2014,Fragile2016}. This result instead argues that our very small disc inner radius is instead the result of modelling systematics such as employing an overly simplified continuum, as discussed above.    

It is important to note that all well-studied QPO models in the literature assume that the QPO frequency increases during the state transition primarily due to the disc inner radius moving inwards \citep{Ingram2019review}. There is therefore no published QPO model that can reproduce our observed QPO frequency for our measured disc inner radius without significant modification.}

\subsection{Misalignment}

\begin{figure}
    \centering
    \includegraphics{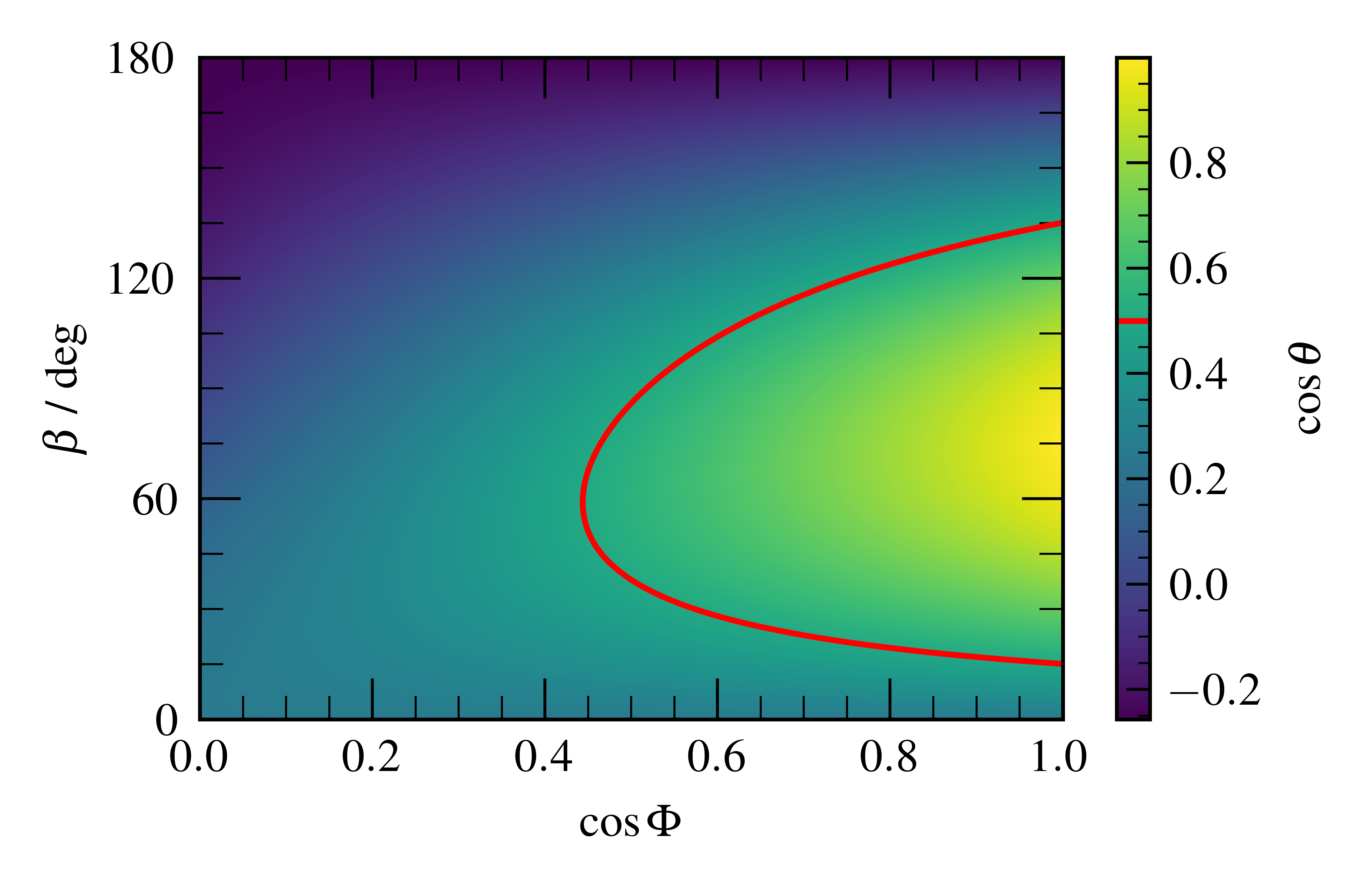}
    \caption{The BH spin axes inclination $\theta$, for different values of misalignment between it and the disc, for the different values of disc azimuthal angle $\Phi$ of the misalignment.  The red line highlights $\theta=60^\circ$, the observed inclination large-scale jet, which could be assumed to be aligned with the BH spin-axis.}
    \label{fig:misalignment_solv}
\end{figure}

Our best fitting disc inclination angle is $i = 75.1^{+0.5}_{-0.3}$ degrees, whereas the jet inclination angle, inferred from radio observations of superluminal ejections \citep{Fender1999}, is $\theta \approx 60^\circ$ \citep[using the radio parallax distance of $D \approx 8.6$~kpc;][]{Reid2014}. We therefore infer a misaligned system, consistent with QPO models that invoke Lense-Thirring precession \citep{stella1998,Ingram2009}. Following I17, we can estimate the misalignment angle $\beta$ between the disc and BH spin axes by taking the large-scale jet as a proxy for the BH spin axis\footnote{This of course might not be correct, as jets have been observed to precess as in e,g, \citet{Miller2019}.}. This is not necessarily simply given by $\beta = i - \theta$, since the azimuthal disc viewing angle, $\Phi$, is unknown. Rather, it can be found by solving the equation \citep{Veledina2013,Ingram2017h1743}
\begin{equation}
    \cos\theta = \sin{i}\sin\beta\cos\Phi + \cos{i}\cos\beta.
    \label{eqn:misalignment}
\end{equation}
Fig.~\ref{fig:misalignment_solv} shows all of the solutions for $\beta$ for the full range of $\Phi$ and $\theta$ values (colour scale), and assuming $i=75^\circ$. The red line represents the solutions corresponding to $\theta=60^\circ$, which cover the range $15.1^\circ<\beta<135.1^\circ$ for $|\Phi|<63.7^\circ$ (i.e. for some values of $\Phi$ there is no solution). The minimum misalginment compatible with our results is therefore $\beta \approx 15^\circ$, which is a large enough misalignment to produce the observed $\sim 15\%$ RMS QPO amplitude with a corona precessing around the BH spin axis \citep{Ingram2017h1743}.

This misalignment will greatly influence the BH spin inferred from disc continuum fitting in the soft state. In the most recent such analysis of GRS~1915+105, \citet{mills2021} report a best fit of $a\approx 0.86$ ($r_\text{isco}(a) = r_\text{in}  \approx 2.57~\rg$), with the very high spin required by reflection spectroscopy measurements (including our own) also within uncertainties. However, they assume a completely aligned system, with the binary inclination $\delta$ equal to the disc inclination $i$ equal to the jet inclination $\theta$. Ignoring relativistic corrections, the disc inner radius inferred from disc fitting is $r_\text{in} \propto D / (M \sqrt{\cos i})~\rg$, and so adopting $i=75^\circ$ in place of $i=\theta=60^\circ$ (but still assuming $\delta=\theta$ and $D=8.6$ kpc) would instead give $r_\text{in} \approx 3.63~\rg$. The inferred inner radius is pushed even further out if we set $\delta = i$ as is assumed in the precession model \citep{Ingram2009,Ingram2015polarization}, since the BH mass is related to the binary mass function $f(M)$ as $M \propto f(M) / \sin^3\delta$. For $\delta = i = 75^\circ$ and still assuming $D=8.6$ kpc, the inner radius becomes $r_\text{in} \approx 5~\rg$, which is incompatible with our measured value of $r_\text{in} \approx 1.4~\rg$. If we instead assume that $\delta$ is unknown, we find that for $i=75^\circ$ and $D=8.6$ kpc the disc continuum fitting inner radius is equal to our value if $\delta \approx 39^\circ$, implying a BH mass of $M \approx 32~\text{M}_\odot$ which is much greater than the dynamical measurement of \chen{$10.1\pm0.6~\text{M}_\odot$ \citep{Steeghs2013}}.

We note, however, that a thin misaligned disc is expected to form a so-called Bardeen-Petterson configuration \citep{bardeen1975}, whereby the outer and inner regions align respectively with the binary and BH spin axes. GRMHD simulations show that the transition from misaligned to aligned can be close to the BH ($r \sim 6$~$\rg$; \citealt{liska2019}), but still further out than our very small inner radius of $r_\text{in} \sim 1.4$~$\rg$. In contrast, our simplified model only considers a planar disc. It could be that failing to account for a more realistic warped disc geometry has introduced a bias into our measurement of $i$, or even of $r_\text{in}$.

\subsection{Biases in the phase-averaged spectrum}

\begin{figure}
    \centering
    \includegraphics{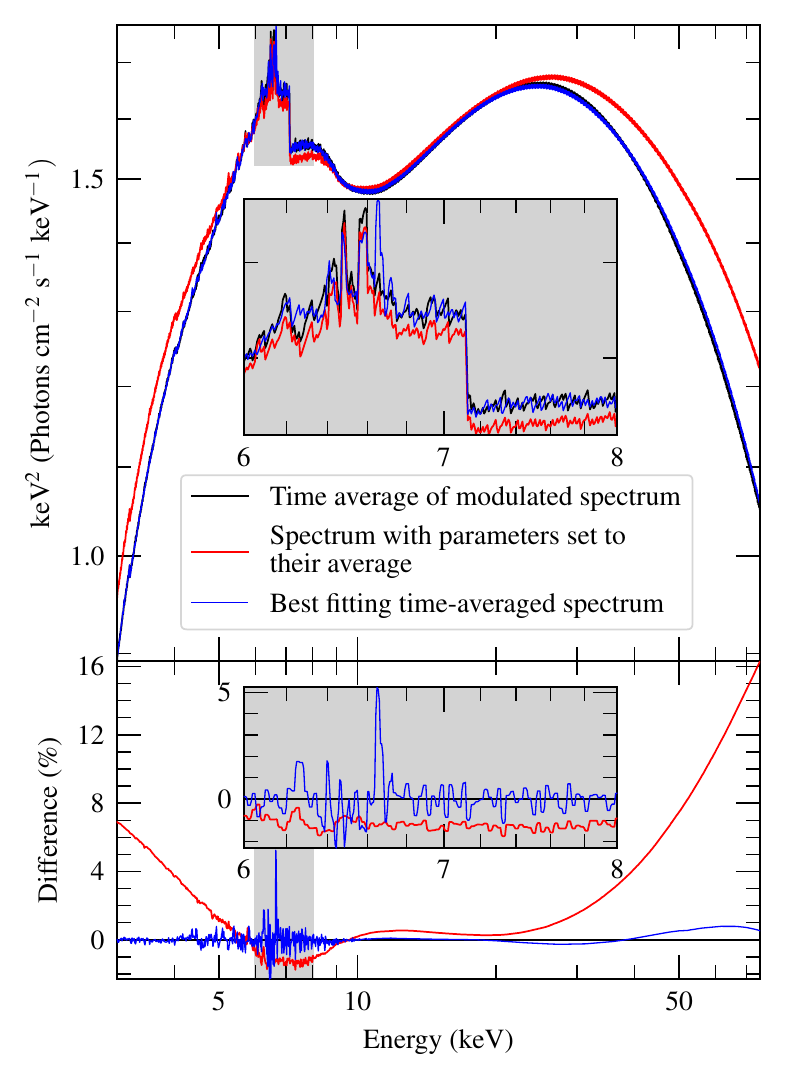}   
    \caption{Top: The best fitting flux-energy specturm (black); the same however with all the modulations turned off so each parameter is set to its phase-average value (red); the best-fitting model when there is no phase-dependant parameter modulations or changing illumination profile (blue).
    Bottom: The percentage difference between the two.  The grey inset panels show a 5-10~keV zoomed-in version of their respective lines, with the corresponding areas of the main graph also highlighted in grey.}
    \label{fig:auto_temp_spec_diff}
\end{figure}

The shape of our best fitting phase-resolved spectrum changes dramatically with QPO phase (Fig.~\ref{fig:nuprec_spectra}). These non-linear variations over each QPO cycle may cause biases in analyses that only fit a steady-state model to the time-averaged flux-energy spectrum. Following I17, we investigate these potential biases in Fig.~\ref{fig:auto_temp_spec_diff} by plotting the phase-averaged spectrum of our best fitting phase-resolved model (black) alongside the spectrum calculated by setting all modulated parameters to their phase-averaged values (red), as well as the percentage difference with respect to the phase-averaged model (bottom panel). We see that the difference between the two spectra is only $\sim 1\%$ in the $\sim 5-20$~keV region, but is much larger at lower and higher energies, indicating that ignoring spectral variability does indeed cause a bias. We investigate further by fitting the observed flux-energy spectrum with a steady-state model (blue lines), which reproduces the phase-averaged model very well (except for a narrow feature at $\sim 6.7$ keV, which \textit{NICER} and \textit{NuSTAR} cannot resolve, but future missions such as \textit{ATHENA} will be able to). The parameters of the new fit are consistent with those of the phase-resolved model except $kT_\text{irr}$, $kT_\text{visc}$ and $kT_e$ are all cooler and the disc normalisation is larger. We therefore conclude that these parameters are biased by spectral variability, but not other key parameters such as $i$ and $r_\text{in}$.

\subsection{Thermalisation lags}

In our model we include a phase lag, $\Delta \gamma$, between variations in the illuminating flux and corresponding variations in disc heating. This is to allow for the finite time it takes for incident radiation to thermalise in the disc atmosphere. This timescale is poorly understood theoretically, but is surely longer than the corresponding timescales for other processes such as fluorescence and scattering, which are almost instantaneous \citep{Garcia2013a}. Leaving $\Delta \gamma$ as a free parameter in principle enables us to empirically measure this thermalisation timescale. Our best fitting value of $\Delta \gamma = 0.15^{+0.06}_{-0.05}$ cycles corresponds to a time lag of $\sim 70^{+26}_{-25}$~ms given that the QPO frequency is $2.2$~Hz, which is much larger than expected. This very long thermalisation time delay is incompatible with observations of $\lesssim 1$~ms soft lags in e.g. MAXI~J~1820+070 \citep{kara2019} and GX~339-4 \citep{Uttley2011}, and so it is very likely that the $\Delta\gamma$ parameter is accounting for some other over-simplification in our model. This is perhaps not surprising, since our prescription is so simple. For instance, we assume a single thermalisation delay, whereas in reality it is likely dependent on disc radius \citep{Frank2002} and on ionisation parameter (which varies with both disc radius and azimuth in our model). Moreover, in reality there will be some smearing such that e.g. very fast fluctuations in the irradiating flux will not be efficiently transferred into fluctuations in disc temperature, which we currently completely ignore.

In the context of the precessing corona model, there are a number of effects that we do not account for here which would lead to modulations of the observed disc temperature. For instance shadowing: the precessing corona and/or jet obscuring different disc azimuths at different precession phases leading to a variation in the shape of the overall observed disc spectrum. We also note that in the precessing corona model, the misalignment between the disc and BH spin axes, $\beta$, remains constant but as the corona precesses around the BH spin axis its misalignment with the disc axis varies between $0$ and $2\beta$ \citep[e.g.][]{Ingram2015polarization}. This would lead to the illumination profile being axisymmetric once per QPO cycle, which is not accounted for by the illumination profile employed here. There are also potential sources of systematic error in our spectral model. For instance in the \textsc{xillverCp} grids that we use, the seed photon temperature is hardwired to $kT_\text{bb}=0.05$ keV, whereas the seed photon temperature of the \textsc{nthcomp} component in our model varies with QPO phase. Finally, we ignore light crossing delays in our parameterisation of the disc illumination profile, which can be the order of milliseconds for reflection from $r \sim 10$s of $\rg$. This is $\lesssim 1\%$ of the $\sim 0.45$s QPO period, but will become increasingly important for the study of QPOs with increasingly higher centroid frequency.

\section{Conclusions}
\label{sec:conclusions}

We have conducted a phase-resolved spectral analysis of a Type-C QPO in a simultaneous \textit{NICER} and \textit{NuSTAR} observation of GRS~1915+105. 
We used a QPO phase-resolving technique, following \citet{Ingram2016} but significantly improved on their work, by including a QPO frequency tracking algorithm that makes it possible to analyse long observations over which the QPO frequency may change, and using the bispectrum to constrain the phase difference between QPO harmonics which has the advantage of enabling a proper Poisson noise correction. We have also used a significantly more advanced version of the \cite{Ingram2017h1743} tomographic model with which to fit the phase-resolved spectra in the Fourier domain. Our new model allows more parameters to be simultaneously modulated and includes self-consistent modulations to the ionisation parameter and disc heating due to irradiation.

Our fit requires the asymmetric illumination parameters $A_1$ and $A_2$ to be $>0$ with $>2\sigma$ confidence. 
Similar to the results of I17 for H~1743-322, this is 
\chai{consistent with that expected for precession of the illuminator, but our measurement has only}
moderate statistical significance.
We also detect a $>3\sigma$ significant modulation of the reflection fraction, indicating that the geometry of the inner accretion disc changes systematically with QPO phase.
We inferred a high disc inclination ($i\approx 75^\circ$), which implies that the disc is misaligned with the previously observed jet ejections ($\theta\approx 60^\circ$). This is consistent with the precessing corona model for the QPO \citep{Ingram2009}. However, our fit also favoured a small disc inner radius, \chai{which requires the corona to be vertically rather than radially extended and is inconsistent with the precession frequency being set purely by the frame dragging effect. We discussed some possible effects that could slow down Lense-Thirring precession compared with the simplest prediction. We note that no QPO model currently in the published literature can reproduce the observed QPO frequency for our measured disc inner radius \citep{Ingram2019review}. Therefore, either our measured radius is affected by modelling systematics -- such as the continuum in reality being more complex than we assume -- or new or modified theories must be developed to model the QPO frequencies.}

We also recovered a large thermalisation delay, which implies that irradiating photons take $\sim 70$ ms to thermalise in the disc atmosphere. We argued that this is infeasibly large and discussed potential sources of systematic error that could be contributing to the measurement.

We compared this analysis and the work of \citet{Ingram2016, Ingram2017h1743} on H~1743-332, and in particular note they both show a strong second harmonic in the modulation of the iron line centroid energy $E_\text{c}(\gamma)$. We found hints that the $E_\text{c}(\gamma)$ modulation evolves systematically with QPO frequency, as would be expected for example if the disc inner radius reduces as the QPO frequency increases. Further work is required to test this hypothesis. 

\section*{Acknowledgements}

A.I., E.N, and D.A acknowledge support from the Royal Society.
MK thanks the support from NWO Spinoza.
We acknowledge the following packages for Python3 \citep{python3_2009} which were used in this analysis:
\begin{itemize}
    \item NumPy \citep{numpy2020}
    \item pandas \citep{pandas2010, pandas2020}
    \item SciPy \citep{scipy2020}
    \item AstroPy \citep{astropy2013, astropy2018}
    \item Matplotlib \citep{matplotlib2007}
    \item Corner \citep{corner2016}
\end{itemize}

This research has made use of data, software and/or web tools obtained from the High Energy Astrophysics Science Archive Research Center (HEASARC), a service of the Astrophysics Science Division at NASA/GSFC and of the Smithsonian Astrophysical Observatory's High Energy Astrophysics Division.

We thank the anonymous referee for their insightful comments.

\section*{Data Availability}

The observational data analysed here are all publicly available via HEASARC. We will share our model upon reasonable request to the corresponding author.




\bibliographystyle{mnras}
\bibliography{bib} 




\appendix

\section{Phase difference from Bispectrum}
\subsection{Treatment of Poisson noise in the Bispectrum}
\label{app:bispec_poisson_noise}
The `jellyfish plots' in
Fig.~\ref{fig:jellyfish} are corrected for Poisson noise and deadtime effects. Failing to correct for these effects introduces an instrumental component into the real part of the bispectrum, which presents itself as a general `drift' of the random walks along the positive real axis. In turn, this reduces the measured bi-phase from the true value.

There exists a Poisson-noise free estimate of the bispectrum, as discussed in \citet{Wirnitzer1985}, which we use in the form
\begin{equation}
    B_m(\nu) = 
    \begin{aligned}[t]
        &R_m(\nu)R_m(\nu)R_m^*(2\nu) \\
        &- 2\lvert R_m(\nu)R^*_m(\nu)\rvert^2 \\
        &-\lvert R_m(2\nu)R^*_m(2\nu)\rvert^2 \\
        &+ 2N\, ,
    \end{aligned}
    \label{eqn:autobispectrum_pn}
\end{equation}
where $R_m(\nu)$ are the Fourier transforms of segments of the lightcurve, and $N = \langle R_m(0)\rangle$ is the average number of photons per light curve segment\footnote{As here for normalisation consistency, we consider light curve segments measured in photon counts, as opposed to the count-rate.}.
Likewise, we want to use a Poisson-free estimate for the normalisation of $b^2(\nu)$.  Considering the expectation of random variables $X$ representing the Fourier transform of a series containing a signal with power $P_\text{s}$, but also noise $P_\text{n}$
\begin{equation}
    \text{E}\left[ \lvert XX \rvert^2\right] = 2 ( P_\text{s} + P_\text{n})^2\,,
    \label{eqn:xx_noise_expectation}
\end{equation}
and clearly in the noiseless case it is $2P^2_\text{s}$.  While we could use a rearrangement of Eq.~\ref{eqn:xx_noise_expectation}, we instead opt to to simply use a noise-subtracted estimate of $P_\text{s}$ from the powerspecturm.  Therefore, we also replace the denominator of Eq.~\ref{eqn:squared_bicoherence} with
\begin{equation}
    2\left(\langle|R_m(\nu)|^2\rangle-N\right)^2 \left(\langle|R_m(2\nu)|^2\rangle-N\right) \,.
    \label{eqn:bispec_normalisation_pn}
\end{equation}

However, the deadtime effects of \textit{NuSTAR} are still more complex than the base Poisson noise.  Therefore we take inspiration from \citet{Bachetti2015} and use the co-bispectrum between light curve segments from FPMA and FPMB ($R_{\text{A},m}(\nu)$ and $R_{\text{B},m}(\nu)$  respectively)
\begin{equation}
    B_m(\nu) = R_{\text{A},m}(\nu)X_{\text{A},m}(\nu)R_{\text{B},m}^*(2\nu)\,,
    \label{eqn:auto-co-bispectrum_pn}
\end{equation}
and also normalise $b^2$ using the denominator
\begin{equation}
        \mathbb{R}\left[\langle R_{\text{A},m}(\nu)R_{\text{B},m}(\nu)\rangle\right]^2 \mathbb{R}\left[\langle R_{\text{A},m}(2\nu)R_{\text{B},m}(2\nu)\rangle\right] \frac{\langle R_{\text{A},m}(0)\rangle}{\langle R_{\text{B},m}(0)\rangle}\,.
    \label{eqn:bispec_normalisation_NuSTAR}
\end{equation}

\subsection{Jellyfish Bootstrapping}
\label{app:bootstrap}
We use a bootstrapping method to calculate the uncertainties in the bispectrum for our determination of the phase-difference between the QPO harmonics.

Each observation has already been split up into $M$ segments, so we draw from this (with replacement) a sample of $M$ segments, $1000$ times\footnote{As an example, if $M=5$, with segments labelled `A',`B',`C',`D', and `E' we would use $1000$ random draws such as `ABAAD', `EECDC', and so on.}. Therefore, each of these $1000$ random draws is the same `length' as the original observation.  We then calculate the bispectrum for each of these samples.  When performing the QPO tracking, we ensure to use the QPO frequency corresponding to the segment's true time.

As $b^2$ is restricted to be positive\footnote{However in the region of low signal/noise, the Poisson correction we make can push an estimate below 0, see Appendix~\ref{app:bootstrap} for details.}, we show the range corresponding to the $\pm1\sigma$ and $\pm3\sigma$ quantiles in Fig.~\ref{fig:jellyfish}.

To ensure we don't fall foul of any issues relating to the cyclic nature of the biphase, to calculate the error on the QPO phase-difference we take the measurement from each $1000$-segment sample.  For these values, we phase-wrap them so each lies within $\pi$~rad of the true measurement and then report the standard deviation on these values as our uncertainty. 

\section{Model Details}
\subsection{Ray tracing}
\label{app:model_raytracing}

A disc patch at coordinate $(r,\phi)$ with radial and azimuthal extent $dr$ and $d\phi$ will subtend a solid angle $d\Omega=d\alpha d\beta/D^2$ on the image plane, and will be centered at horizontal and vertical coordinates on the image plane $\alpha$ and $\beta$. Here, $D$ is the distance from observer to source, and $\alpha$ and $\beta$ are the impact parameters at infinity, where the singularity occupies the position $\alpha=\beta=0$ on the image plane. The total QPO phase-dependent specific flux observed from the disc is (e.g. \citealt{Ingram2019reltrans})
\begin{equation}
    F_d(E,\gamma) = \int_0^{2\pi} \int_{r_\text{in}}^{r_\text{out}} g^3(r,\phi) \frac{I(E/g,r,\phi,\gamma)}{D^2} d\alpha d\beta,
    \label{eqn:Fd}
\end{equation}
where $g=E/E_d$ is the energy shift experienced by a photon travelling from disc coordinate $(r,\phi)$ to the observer (given by Equation 4 in \citealt{Ingram2017h1743}) and $E$ is photon energy in the observer restframe.

\subsection{Model Normalisation}
\label{app:model_norms}

We normalise the \textsc{xillverCp} spectrum, $\mathcal{R}(E)$, such that the incident spectrum that goes into the calculation has an integral over all energies of unity. Due to the internal \textsc{xillverCp} normalisation (see Eq.~16 of \citealt{Ingram2019reltrans}), this gives
\begin{equation}
    \int_0^{2\pi} \int_0^1 \int_0^\infty \mu_e \mathcal{R}(E) dE d\mu_e d\phi = 1,
    \label{eqn:Rnorm}
\end{equation}
where $\mu_e$ is the cosine of the emission angle. We normalise the emissivity function such that
\begin{equation}
    \int_{\alpha,\beta} g^4(r,\phi) \epsilon(r,\phi,\gamma) d\alpha d\beta= 1.
    \label{eqn:epsnorm}
\end{equation}
These two conditions together ensure that the observed bolometric reflected flux (in the case of $N_\text{d}=0$) would be $f_\text{R}(\gamma) N_\text{c}(\gamma)$ \textit{if} the function $\mathcal{R}(E)$ was independent of emission angle.

\section{Phase offsets with different reference bands}
\label{app:phi_c}
Recalling Eq.~\ref{eqn:FT_spectra}, the FT of the $j^\text{th}$ QPO harmonic is
\begin{equation}
    W_j(E) = \mu(E) \sigma_j(E) \text{e}^{\text{i}\Phi_j(E)},
\end{equation}
where (recalling Eq.~\ref{eqn:phase_offsets})
\begin{equation}
   \begin{split}
        \Phi_1(E) &= \Phi_1 + \Delta_1(E) \\
        \Phi_2(E) &= 2(\Phi_1 +\psi) + \Delta_2(E).
    \end{split}
    \label{eqn:appendix_phase_offsets}
\end{equation}

Here, denoting the subject band as $S_j$ and reference band as $R_j$, the phase lags are (taking the argument of Eq.~\ref{eqn:cross_spectrum})
\begin{equation}
    \Delta_j(E) = \text{arg}\left[  S_j(E) R_j^*  \right].
\end{equation}

When we measure the same QPO FT but with a different instrument, and therefore with a different reference band, we consider a different reference band $T$, where $T_j$ lags behind $R_j$ by a phase difference $\delta_j$. The QPO FT we measure from our new instrument is therefore
\begin{equation}
    Q_j^T(E) = \mu^T(E) \sigma_j(E) \text{e}^{\Phi_j^T(E)},
\end{equation}
where $\Phi_j^T(E)$ is
\begin{equation}
    \begin{split}
        \Phi_1^T(E) &= \Phi_1 + \Delta_1(E) - \delta_1 \\
        \Phi_2^T(E) &= 2 ( \Phi_1 + \psi^T ) + \Delta_2(E) - \delta_2,
    \end{split}
    \label{eqn:t_band_phase_offsets}
\end{equation}
and where $\psi^T$ is the phase difference between harmonics in the $T$ band, which in principle can be different from $\psi$ as the reference bands come from instruments with different energy bands, and even have different responses within overlapping bands.

Combining Eqs.~\ref{eqn:appendix_phase_offsets} and \ref{eqn:t_band_phase_offsets}, we find that 
\begin{equation}
    \begin{split}
        \Phi_1^T(E) &= \Phi_1(E) - \delta_1 \\
        \Phi_2^T(E) &= \Phi_2(E) + 2 ( \psi^T - \psi ) - \delta_2\,,
    \end{split}
\end{equation}

Considering the phase difference between the harmonics in any energy band \citep{Ingram2016}
\begin{equation}
    \psi(E) = \psi - \Delta_1(E) + \frac12 \Delta_2(E).
\end{equation}
we can similarly construct that
\begin{equation}
    \psi^T = \psi - \delta_1 + \frac12 \delta_2\,,
\end{equation}
and therefore
\begin{equation}
    \begin{split}
        \Phi_1^T(E) &= \Phi_1(E) - \delta_1 \\
        \Phi_2^T(E) &= \Phi_2(E) - 2\delta_1.
    \end{split}
\end{equation}

Putting this together, the relation between the QPO FT when measured with the different reference bands is
\begin{equation}
    Q_j^T(E) = Q_j(E) \exp( -\text{i} j \delta_1  ),
\end{equation}
and so the phase offset of the $j^\text{th}$ harmonic is
\begin{equation}
    \phi_\text{c, j} = -j \delta_1,
\end{equation}
so finally we have
\begin{equation}
    \begin{split}
        \phi_\text{c,1} &\equiv \phi_\text{c} \\
        \phi_\text{c,2} &= 2\phi_\text{c}.
    \end{split}
\end{equation}

\section{MCMC corner plots of modulated parameters}
\label{app:param_corner_plots}
\begin{figure*}
    \centering
    \includegraphics{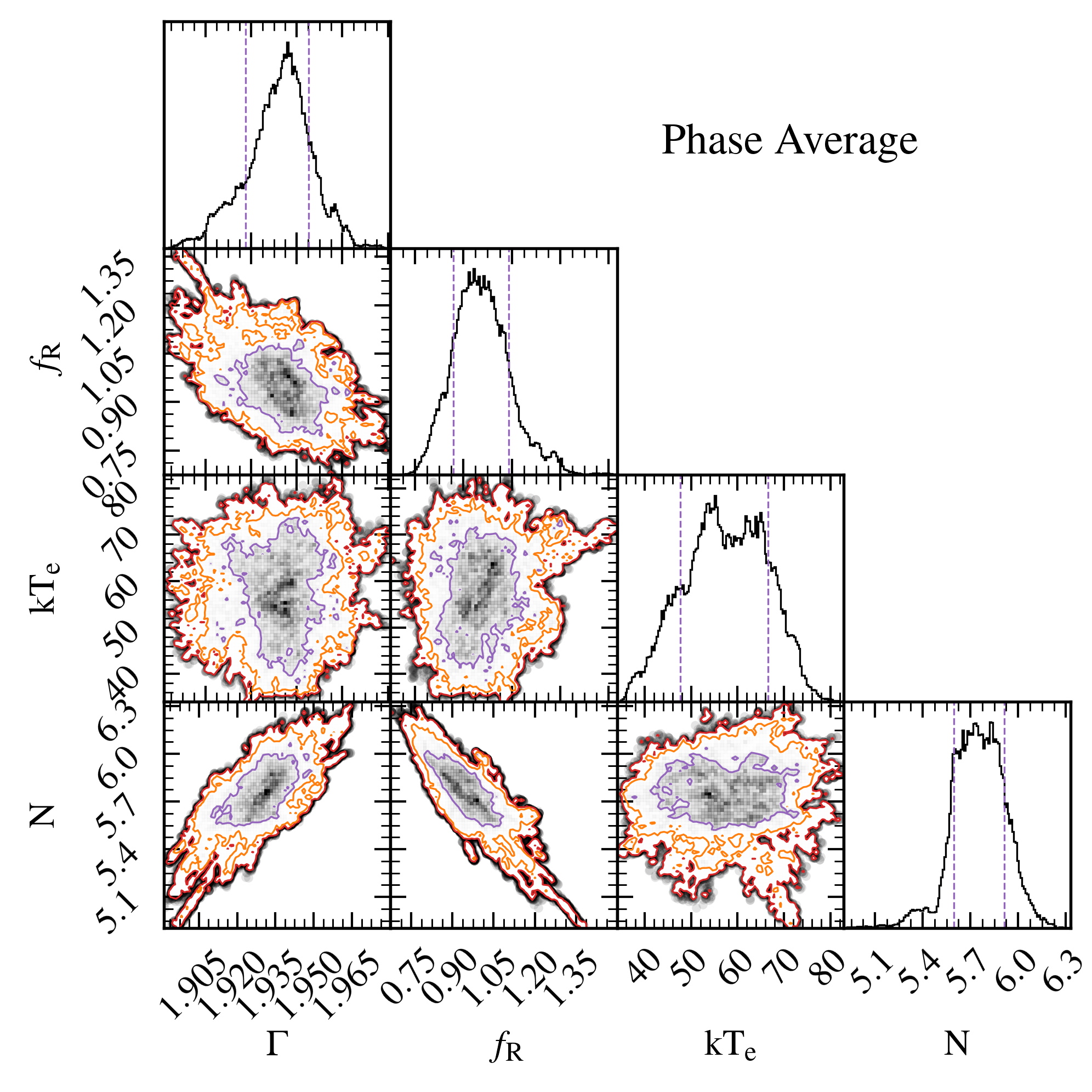}
    \includegraphics{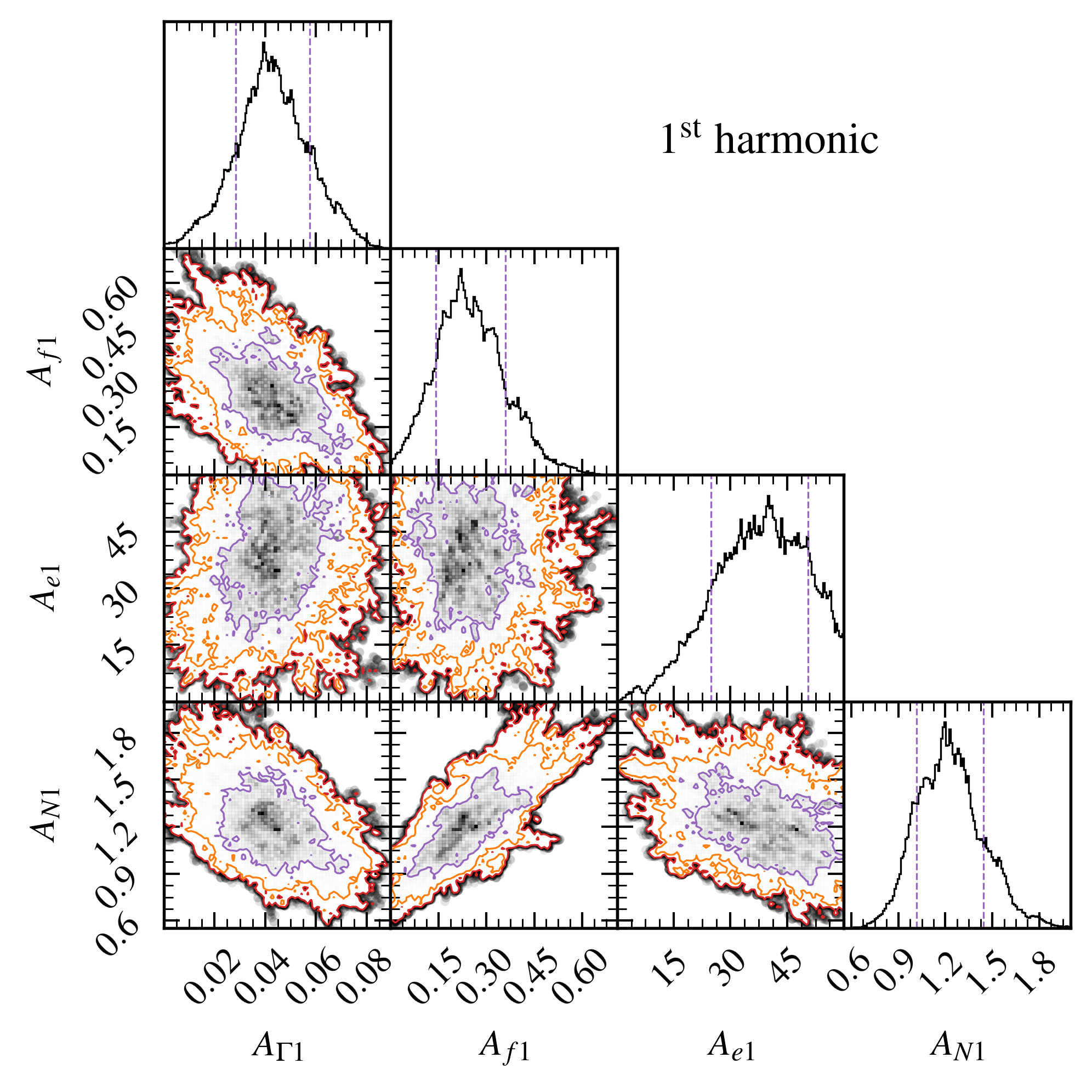}
    \includegraphics{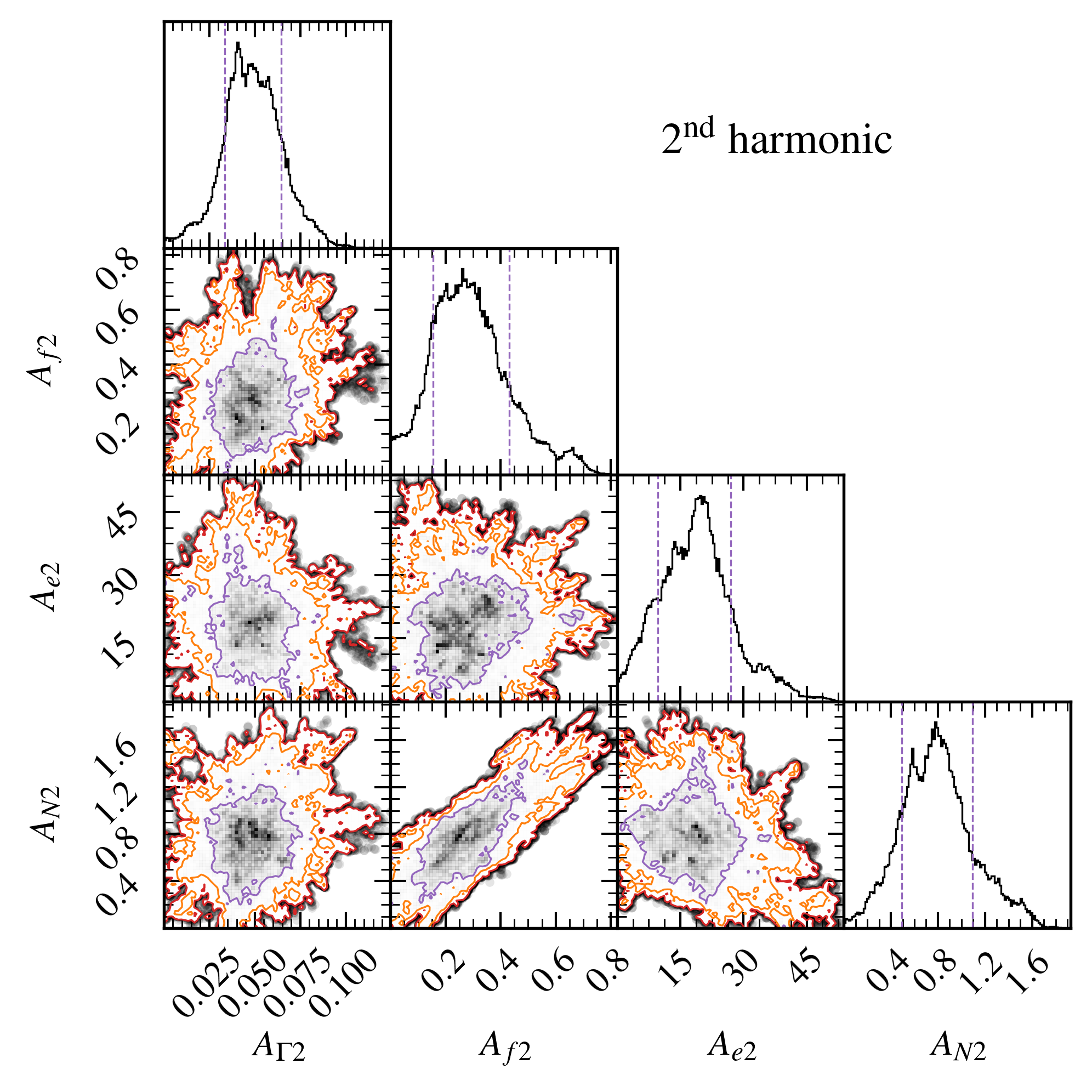}
    \caption{Corner plots from the MCMC of the modulated parameters within the model.  The purple, orange, and red lines show the $1,2,3\sigma$ credible intervals respectively. The phase-average, and also the amplitudes of the first and second harmonics are shown separately.}
    \label{fig:param_harmonics_corner_plots}
\end{figure*}

\begin{figure*}
    \centering
    \includegraphics{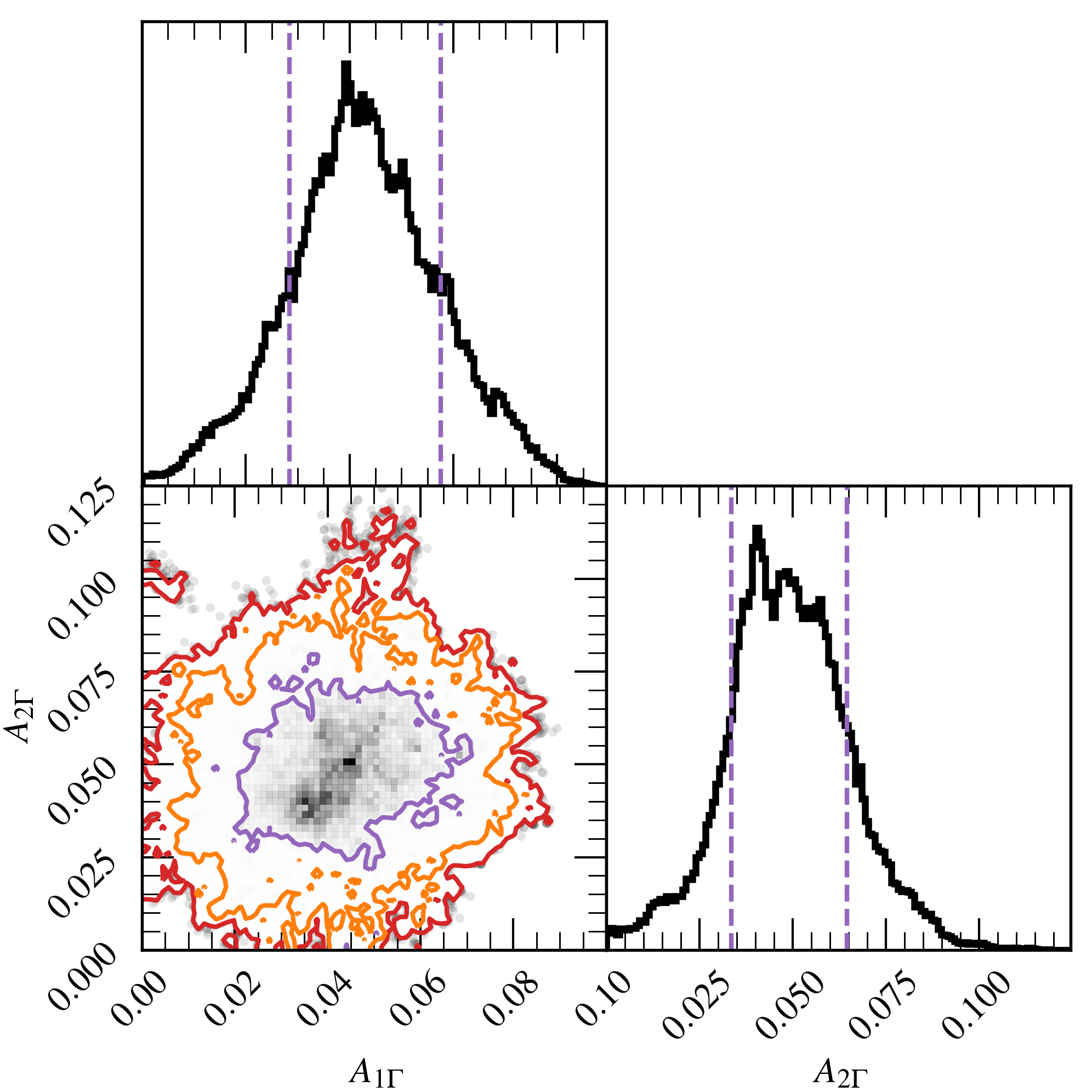}
    \includegraphics{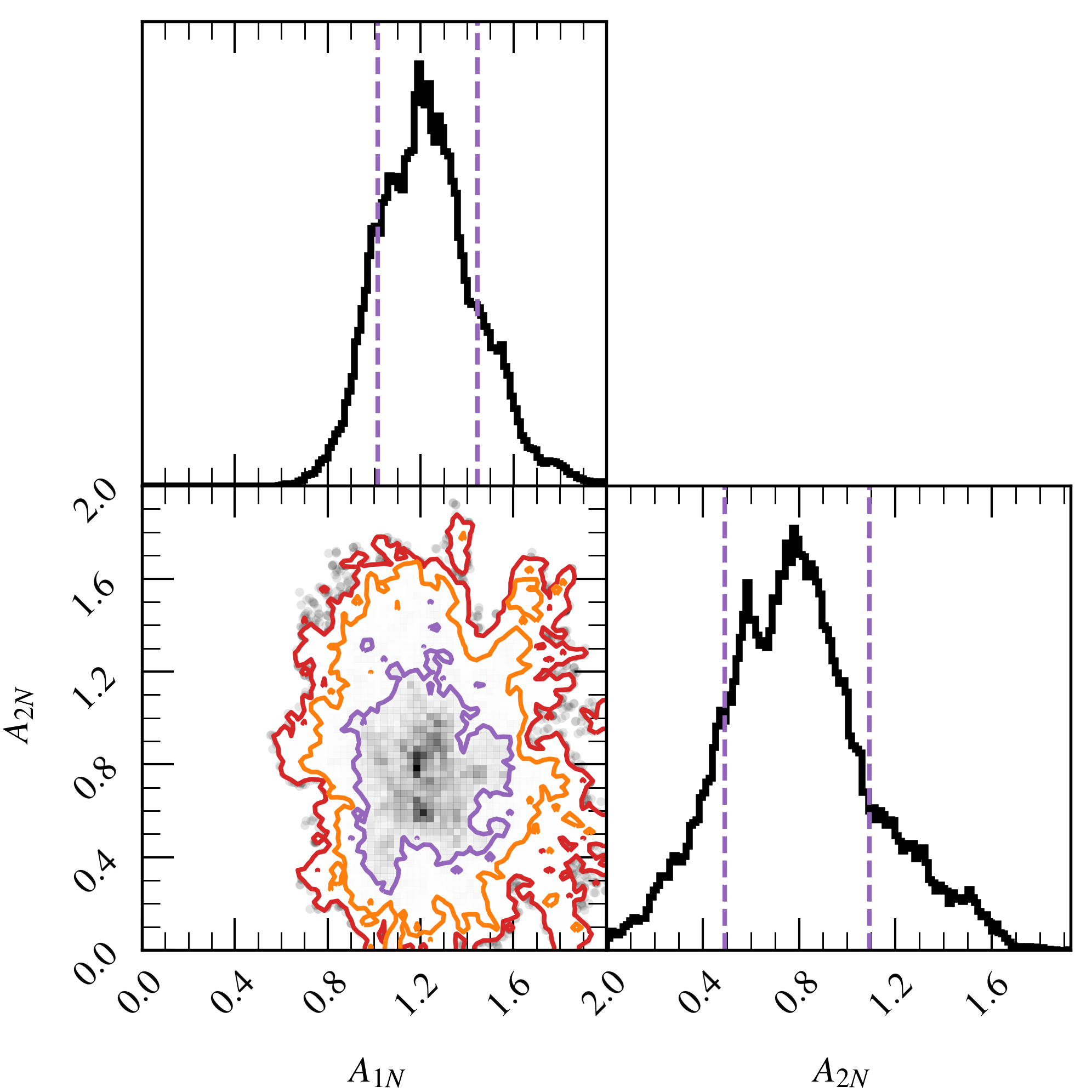}
    \includegraphics{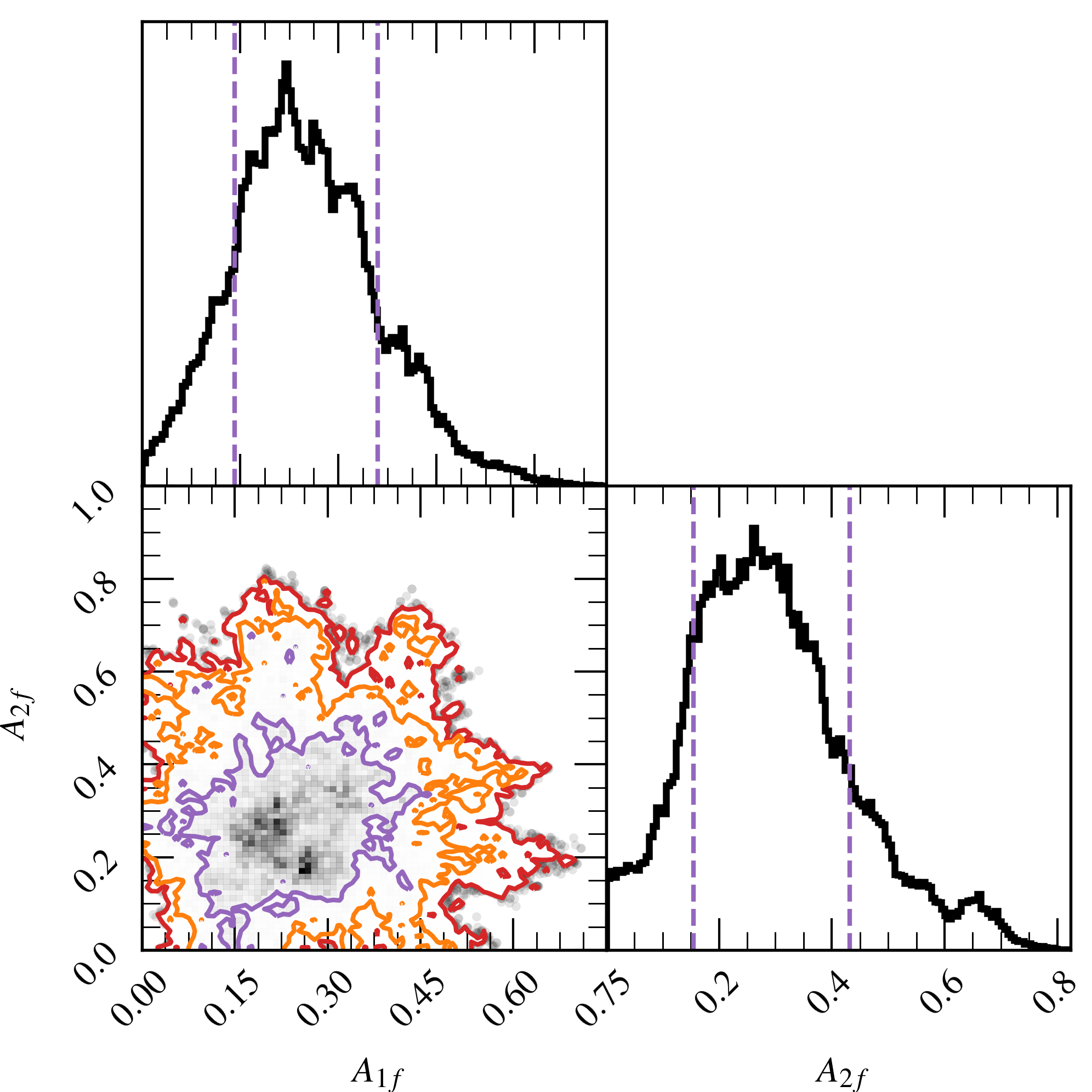}
    \includegraphics{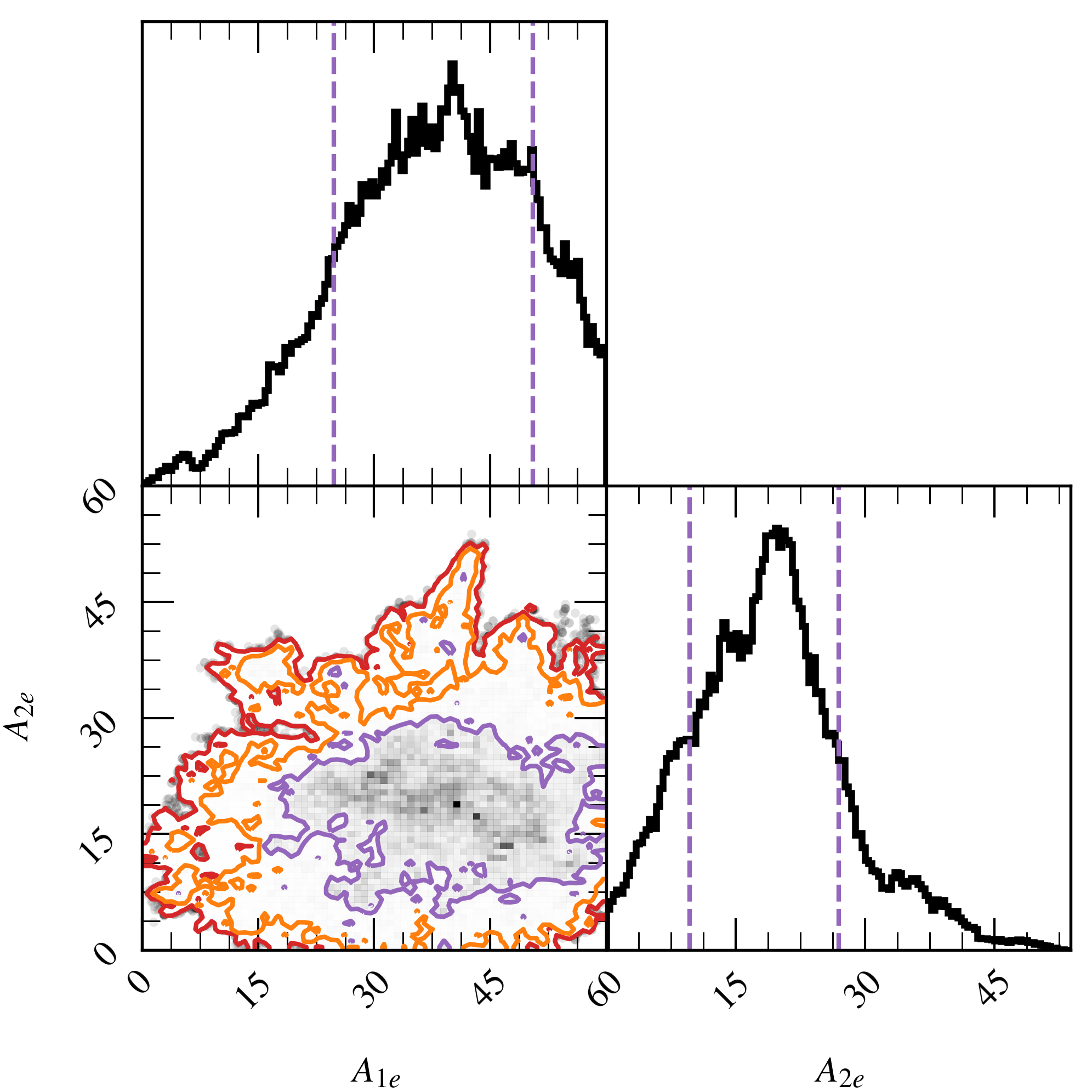}
    \caption{Corner plots from the MCMC of the amplitudes of the modulations  within the model.  The purple, orange, and red lines show the $1,2,3\sigma$ credible intervals respectively.  In all cases, the $(0,0)$ lies outside of the $3\sigma$ contour showing that all four parameters are consistent with being modulated.}
    \label{fig:param_modulation_corner_plots}
\end{figure*}

This appendix contains corner plots of some of the parameters from the MCMC.  Fig.~\ref{fig:param_harmonics_corner_plots} shows the phase average, and the $1^\text{st}$ harmonic and $2^\text{nd}$ harmonic amplitudes of the modulated parameters compared with each other.  Fig.~\ref{fig:param_modulation_corner_plots} shows the $1^\text{st}$ and $2^\text{nd}$ harmonic amplitudes for each of the modulated parameters, from which we can see how significantly away from $(0,0)$ the parameter is, which would correspond to that parameter not being modulated.

\begin{figure*}
    \centering
    \includegraphics{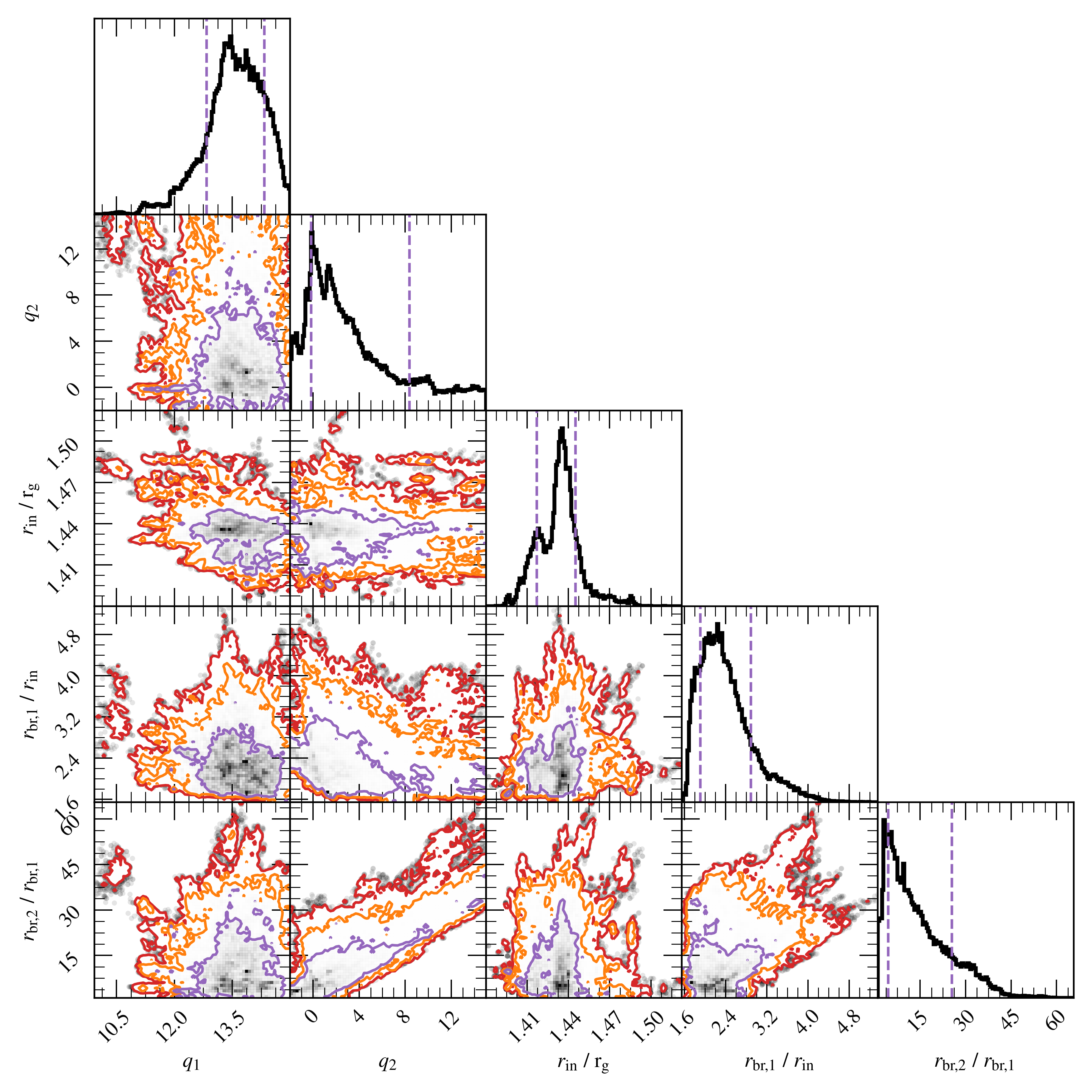}
    \caption{Corner plots from the MCMC of the radial emissivity parameters.  The purple, orange, and red lines show the $1,2,3\sigma$ credible intervals respectively.  It is important to note that the two break radii $r_\text{br,1}$ and $r_\text{br,2}$ have units of the disc truncation radius $r_\text{in}$ and $r_\text{br,1}$ respectively.  As $q_1$ and $q_2$ are the power law indices within there relevant break radii, the model becomes insensitive to them when $r_\text{br,1}$ and $r_\text{br,2}$ approach $1$ respectively.}
    \label{fig:radial_emissivity_corner_plot}
\end{figure*}

\begin{figure*}
    \centering
    \includegraphics{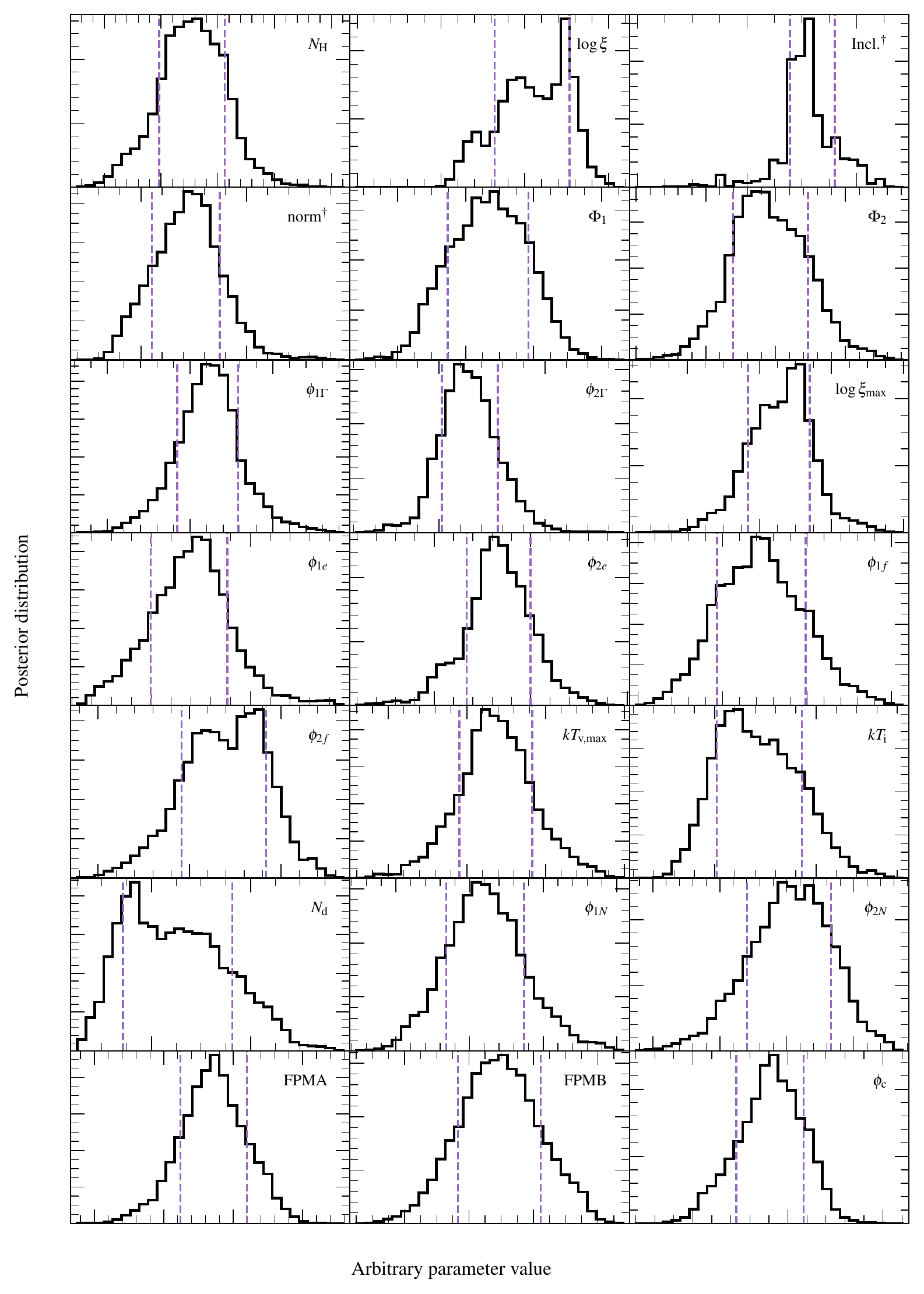}
    \caption{The posteriors of all the parameters not otherwise shown in this paper, with the $\pm1\sigma$ credible interval shown by the dashed purple vertical lines.  For visual simplicity the parameter values are not shown here, as the shape of the distribution is the key feature.  The parameters marked with $^\dagger$ relate solely to the distant reflector component of the model.}
    \label{fig:all_other_param_hists}
\end{figure*}

For the majority of the parameters, the posterior is mostly symmetric and approximately Gaussian.  However, this is not true for some of radially emissivity parameters $q_1$, $q_2$, $r_\text{in}$, $r_\text{br,1}$, and $r_\text{br,2}$ likely due to their interdependence (see Eq.~\ref{eqn:eps2}).  For completeness they are shown as a corner plot in Fig.~\ref{fig:radial_emissivity_corner_plot}; all parameters whose posterior distributions are not otherwise shown in this paper are included as histograms in Fig.~\ref{fig:all_other_param_hists} with symbols matching those in Table~\ref{tab:auto_temp_free_pars}.

\bsp	
\label{lastpage}
\end{document}